\documentclass[12pt]{article}
\def\bea{\begin{eqnarray}}
\def\eea{\end{eqnarray}}
\def\be{\begin{equation}}
\def\ee{\end{equation}}
\def\re#1{(\ref{#1})}

\begin{document}

\begin{titlepage}
\vskip 1. cm
\begin{center}
\huge{\bf{Kaluza-Klein towers for spinors in warped spaces}}
\end{center}
\vskip 0.6 cm
\begin{center} {\Large{Fernand Grard$^1$}},
\ \ {\Large{Jean Nuyts$^2$}}
\end{center}
\vskip 1 cm

\noindent{\bf Abstract}
\vskip 0.2 cm
{\small
\noindent
All the boundary conditions compatible with the reduction of a five dimensional spinor field
of bulk mass $M$ in a compactified warped space to a four dimensional brane are 
derived from
the hermiticity conditions of the relevant operator. The possible presence of metric singularities
is taken into account. Examples of resulting Kaluza-Klein spinor towers are given for a representative set of
values for the basic parameters of the model and of the parameters describing the allowed boundary 
conditions, within the hypothesis that there exists one-mass-scale-only, the Planck mass. In many cases, the 
lowest mass in the tower is small and very sensitive to the parameters while the other masses 
are much higher and become more regularly spaced.
In these cases, if a basic fermion of the standard model (lepton or quark) happens 
to be the
lowest mass of a Kaluza-Klein tower, the other masses would be much larger 
and weakly dependent on the fermion which defines the tower.  
 }
\vfill
\noindent
{\it $^1$  Fernand.Grard@umh.ac.be,  Physique G\'en\'erale et
Physique des Particules El\'ementaires,
Universit\'e de Mons-Hainaut, 20 Place du Parc, 7000 Mons, Belgium}
\vskip 0.2 cm
\noindent
{\it $^2$  Jean.Nuyts@umh.ac.be,
Physique Th\'eorique et Math\'ematique,
Universit\'e de Mons-Hainaut,
20 Place du Parc, 7000 Mons, Belgium}
\end{titlepage}

\newpage
\section{Introduction \label{Intro}}

	In recent articles, we have analysed the generation
of Kaluza-Klein mass towers in five dimensional theories with a fifth dimension
compactified either to a strip or on a circle. This study
was carried out considering a scalar field propagating freely in the bulk, at first
in a flat space \cite{GN1}, then in a warped space \cite{RS1} without metric singularities \cite{GN2}
and finally in a warped space with metric singularities \cite{GN3}. The approach relies
on a careful study of the hermiticity properties of the 
operators which arise in the Kaluza-Klein reduction equations 
and which are of
second order in the derivatives. 
This lead us, considering different five-dimensional metric configurations
(flat and warped),
to classify 
all the sets of allowed boundary conditions and, from them, 
the corresponding mass equations leading to the construction
of the so-called Kaluza-Klein mass towers \cite{KK}.
Remember that the consideration of warp spaces offers the 
possibility to solve in a elegant way,  with only one extradimension,
the hierarchy problem, in the sense that starting from the Planck mass 
as the only fundamental mass of the model, the observable 
low lying Kaluza-Klein masses can be made of the order of TeV without fine-tuning.

Having in mind that the future high energy colliders are expected 
to look for the possible appearance of Kaluza-Klein
mass towers which could be of non zero spin 
as evidence for the existence of fields with spin 
propagating in higher dimensions, we were led to extend our work
to spinor fields. In a previous article \cite{GN4}, 
we restricted ourselves, as a first step in a more general
approach, to a five-dimensional flat space. 
Requesting the Dirac operator be a symmetric operator and taking into account
the underlying symmetries of the Dirac equation in five dimensions, 
in particular covariance and parity invariance 
in the brane, the whole set of allowed boundary conditions 
has been established leading to the mass equations from which 
the Kaluza-Klein mass towers are built.

In our preceding papers, 
illustrative numerical examples of Kaluza-Klein mass towers 
were given for the different configurations we considered.

In this article, we extend our study of Dirac fields by 
the consideration of five dimensional compactified 
warp spaces, first without, and then with metric singularities.     
                                                                                                                                      The article is organized as follows.
In Sect.\re{Diracgen}, we consider the case of a 
warp space with no metric singularity. 
We establish the specific form of the Dirac equation
and proceed with the Kaluza-Klein reduction.
The whole set of allowed boundary conditions 
are obtained from the hermiticity of the Dirac operator.
The solutions for a free field with an arbitrary mass M 
propagating in the bulk are given explicitly. 
In Sect.\re{warpedspaceN}, 
we extend the same considerations to the case of warp spaces
with an arbitrary number of singularities. 
Sect.\re{physdis1} is devoted to various physical considerations
concerning the determination and the interpretation 
of the Kaluza-Klein mass eigenstates, in particular considerations about the possible 
choices of boundary conditions, the closure 
of the extradimension strip to a circle, the mass scales of the model, the 
relation between Kaluza-Klein eigenmasses 
and observable masses and finally the mass state probability densities.
In Sect.\re{towers}, the general procedure 
adopted for the determination of the
Kaluza-Klein mass towers is elaborated from the boundary 
conditions and the analytical expressions of the field.
In Sect.\re{Numeric}, some illustrative 
numerical examples of Kaluza-Klein mass towers are 
given for specific boundary conditions, 
in the cases without and with metric singularities.

In App.\re{leastact}, 
we show that the boundary relations derived from the application of
the least action principle are identical 
to those we deducted from the symmetry of the Dirac operator.
In App.\re{ExempBC}, we developed 
some examples of boundary conditions in the general case of $N$ metric 
singularities.

\newpage

\section{The Dirac equation in a warped space. No metric singularities \label{Diracgen}}

\subsection{The Dirac equation. Invariances \label{Dirac}}

We study the Dirac equation (see App.\re{leastact}) with a bulk mass $M$
\begin{equation}
\left(i\gamma^AD_A-M\right)\Psi=0
\label{Diracgen5}
\end{equation}
and with the invariant scalar product between the spinors $\Phi$ and $\Psi$
\begin{equation}
\left(\Phi,\Psi\right)= \int \overline{\Phi}(x)\Psi(x) \sqrt{g}\, d^4 x 
\label{scalprodspin}
\end{equation} 
in a five dimensional warped space.
With the following notation for the indices
\be
\left\{
   \begin{array}{ccccccc}
{\rm{warp\ space}}&:&
      \left\{A\right\}
     \equiv \left\{{\displaystyle{\Sigma}},{\displaystyle{S}}\right\}
      \equiv\left\{{\displaystyle{0}},{\displaystyle{I}},{\displaystyle{S}}\right\}
        &,& {I}={\displaystyle{1}},{\displaystyle{2}},{\displaystyle{3}}&,&
       {\displaystyle{S}}={\displaystyle{5}}
      \cr
{\rm{local\ space}}&:&
       \left\{a\right\}
       \equiv\left\{\sigma,s\right\}
       \  \equiv \left\{0,i,s\right\}
        &,& i=1,2,3&,&
       s=5 
    \end{array}
    \right.         
\label{notationspace}
\ee
\be
0{\displaystyle{0}}
\ee
the warped metric is
\begin{equation}
dS^2=g_{AB}dx^{A}dx^{B}=
g_{\Sigma\Theta}dx^{\Sigma}dx^{\Theta}-ds^2
     =\lambda^2 e^{-2\epsilon k s}
     \eta_{\sigma\theta}dx^{\Sigma}dx^{\Theta}-ds^2\ .
\label{ds2warp}
\end{equation}
In this equation, $\eta_{\sigma\theta},\eta_{ss}$ are the components of
the five dimensional flat space metric with 
signature $(+,-,-,-,-)$,
$\lambda$ is an arbitrary positive constant, introduced for later convenience,
while, with $k$ defined to be positive, the warp factor $\epsilon k,\ \epsilon=\pm 1$ 
can be chosen to be positive or negative. As in the four-dimensional space,
the Dirac spinor is four-dimensional $\Psi_{\alpha},\ \alpha=1,\dots,4$
in a five dimensional space.

We now compute explicitly the vielbein and the related quantities 
($\gamma$-matrices and covariant derivatives) needed
to put the Dirac equation in the form \re{dirac2warp} suitable for its application to the case
of interest.   

The non zero elements of the vielbein $e_A^{\phantom{D}c}$ defined as usual by
\begin{equation}
g_{AB}=e_A^{\phantom{D}c}\,\eta_{cd}\,e_B^{\phantom{D}d}
\label{vielbeindef}
\end{equation}
are chosen as
\bea
e_{\Sigma}^{\phantom{D}\sigma}
=e_{\displaystyle{0}}^{\phantom{D}0}=e_{\displaystyle{1}}^{\phantom{D}1}
=e_{\displaystyle{2}}^{\phantom{D}2}=e_{\displaystyle{3}}^{\phantom{D}3}
&=&\lambda e^{-\epsilon ks}
\ \quad,\quad e_{S}^{\phantom{D}s}=1
 \label{vielbein1warp}\\
e_{\sigma}^{\phantom{D}\Sigma}
=e_{0}^{\phantom{D}\displaystyle{0}}=e_{1}^{\phantom{D}\displaystyle{1}}
=e_{2}^{\phantom{D}\displaystyle{2}}=e_{3}^{\phantom{D}\displaystyle{3}}
&=&\lambda^{-1} e^{\epsilon ks}
\quad,\quad e_{s}^{\phantom{D}S}=1\ .
\label{vielbein2warp}
\eea
We take local $\gamma^{a}$ as those of the flat five-dimensional space 
($[\gamma_a,\gamma_b]_+=\eta_{ab}$). They can be built from a set 
$\gamma_{\sigma},\ \sigma={0,\dots,3}$ matrices
of the four dimensional flat space. In particular 
one has $\gamma_s=\gamma_0\gamma_1\gamma_2\gamma_3$. 
The warped
$\gamma^{A}$ are given by 
\begin{equation}
\gamma^{A}=e_{a}^{\phantom{D}A}\gamma^{a}\ .
\label{gammawarp}
\end{equation}

The Dirac equation is covariant under the diffeomorphisms 
\bea
   x^{'A}&=&x^{'A}(x^B)
     \nonumber\\
   dx^{'A}&=&\frac{\partial x^{'A}}{\partial x^B}\,dx^B   
     \nonumber\\
   e_{a}^{'\phantom{D}A}(x')&=&\frac{\partial x^{'A}}{\partial x^B}\, e_{a}^{\phantom{D}B}(x)
     \nonumber\\
   \Psi'(x')&=&\Psi(x)  
\label{diffeo}
\eea
and under the local SO(4,1) local transformations $T^{a}_{{\phantom{a}}b}(x)$ ($T^t\eta T=\eta$)
\be
\begin{array}{cclcccl}
      {\widehat{e}}^{\phantom{D}a}_{A}&=&T^{a}_{{\phantom{a}}b}e^{\phantom{D}b}_{A}
      &,& \eta_{ad}&=&T^{b}_{\phantom{A}a}\ \eta_{bc}\ T^{c}_{\phantom{A}d}    
      \nonumber\\
      {\widehat{\Psi}}_{\alpha}&=&S^{[T]}_{\alpha\beta}\Psi_{\beta}&&&&
\end{array}
\label{localso41}
\ee
where $S^{[T]}_{\alpha\beta}$ is the spinor transformation corresponding 
to the vector transformation $T^{a}_{{\phantom{a}}b}$, 
in particular $S^{-1}\gamma^{a}S=T^{a}_{\phantom{A}b}\gamma^{b}$.

The non zero elements of the covariant derivatives 
of the vielbein are computed to be
\bea
&&\left(D_{\displaystyle{0}}e\right)_{\displaystyle{5}}^{\phantom{D}o}
= \left(D_{\displaystyle{I}}e\right)_{\displaystyle{5}}^{\phantom{D}i}
=\epsilon \lambda k e^{-\epsilon ks}
    \nonumber\\
&&\left(D_{\displaystyle{0}}e\right)_{\displaystyle{0}}^{\phantom{D}5}
= -\left(D_{\displaystyle{I}}e\right)_{\displaystyle{I}}^{\phantom{D}5}
=\epsilon \lambda^2 k e^{-2\epsilon ks}\ .    
\label{covdervierwarp}
\eea
The covariant derivative of the four-component spinor field $\Psi_{\alpha}$
is given by
\begin{equation}
\left(D^{[s]}_A\Psi\right)_{\alpha}=\partial_A\Psi_{\alpha}+\left(G^{[s]}_A\right)_{\alpha\beta}\Psi_{\beta}
\label{covderspin5}
\end{equation}
where the spinor connection $\left(G^{[s]}_A\right)_{\alpha\beta}$, written in general
\begin{equation}
G_A^{[s]}=-\frac{i}{2}g^{BC}e_{B}^{\phantom{D}a}\left(D_A e\right)_{C}^{\phantom{D}b}\sigma_{ab}\ ,
\label{Gammafinal}
\end{equation}
reduces here to
\begin{equation}
G^{[s]}_{\Sigma}=
-\frac{1}{2}\epsilon \lambda k e^{-\epsilon k s} \gamma_{\sigma}\gamma^5\ .
\label{connectionwarp}
\end{equation}

Collecting the terms, one finds from \re{Diracgen5}
the specific Dirac equation of the warped space \re{ds2warp}
\begin{equation}
\biggl(
\frac{e^{\epsilon k s}}{\lambda}\bigl(i\gamma^{\sigma}\partial_{\Sigma}\bigr)
+\bigl(i\gamma^{5}\bigr)\left(\partial_{\displaystyle{5}}-2k\epsilon\right)\biggr)\Psi = M \Psi \ .
\label{dirac2warp}
\end{equation}
The five-dimensional mass $M$, the mass in the bulk,
is an arbitrary parameter of the model. 

\subsection{Symmetry of the Dirac operator}

In order to have real $M$, the Dirac operator $\cal{D}$ in \re{Diracgen5}
\begin{equation}
{\cal{D}}=i\gamma^{A}D_{A}
\label{Diracop}
\end{equation}
should be symmetric for the hermitian scalar product \re{scalprodspin}, namely
\begin{equation}
\left(\Phi,{\cal{D}}\Psi\right)=\left({\cal{D}}\Phi,\Psi\right)\ .
\label{symdirac}
\end{equation}
Using the identity 
\begin{equation}
\frac{1}{2} \sqrt{g} \bigl(g^{AB}\eta_{ac}
                 -e_a^{\phantom{B}B}e_c^{\phantom{B}A}\bigr)\left(D_A e\right)_B^{\phantom{B}c}                                 
                        -\partial_A\left(\sqrt{g}e_a^{\phantom{B}A}\right)=0
\label{vielbeinid0}
\end{equation}which can be proved using
\begin{equation}
\partial_A\left(\sqrt{g}\right)=\frac{1}{2}g^{BC}\left(\partial_Ag_{BC}\right)\ ,
\label{derdetg}
\end{equation}
the equation \re{symdirac} reduces (up to a factor $i$) to the integral of a divergence
\begin{equation}
\int\partial_A\biggl({\overline{\Phi}}\gamma^{A} \Psi\sqrt{g} \biggr)\,d^d x=0
\label{integral}
\end{equation}
meaning that this operator is formally symmetric.
This equation determines the boundary conditions which must be satisfied by $\Phi$
and $\Psi$ in order for the Dirac operator to be fully symmetric. In this Section,
the discussion is carried on in a warped space without metric singularities 
and in Section \re{warpedspaceN} with an arbitrary number $N$ of metric singularities. 

\subsection{Kaluza-Klein reduction.
No metric singularity  \label{KKreduc}}

We adopt the following Kaluza-Klein separation of variables
\begin{equation}
\Psi(x^{\mu},s)=\sum_n \bigl(F^{[n]}(s)+iG^{[n]}(s)\gamma^{5}\bigr)\psi^{[n]}(x^{\mu})
\label{warp}
\end{equation}
assuming that $F^{[n]}(s)$ and $G^{[n]}(s)$ are complex functions depending on $s$ only 
while $\psi^{[n]}(x^{\mu})$ is a $x^{\mu}$ dependent spinor.

In this form we have made the most general choice compatible 
with an SO(3,1) spinor invariance in the sense that $\Psi$ and $\psi^{[n]}$ 
are supposed to transform in the same way under this subgroup of the spinor
SO(4,1) transformations \re{localso41}.

\subsection{Boundary relation and conditions for the spinor fields. 
No metric singularity \label{boundrel}}

For each variable $x^{A}$ with range $[-\infty,\infty]$, the integration in \re{integral}
is identically zero for $\Psi$ and $\Phi$ in the spinor Hilbert space (sufficient decrease 
of the fields at $\pm\infty$).
For the variables $s$ which has a finite range $[0,2\pi R]$, the boundary relation is
\begin{equation}
\int d^{4}\!x\ \biggl[{\overline{\Phi}}\gamma^{5} \Psi\sqrt{g} \biggr]_{s=2\pi R}
=\int d^{4}\!x\ \biggl[{\overline{\Phi}}\gamma^{5} \Psi\sqrt{g} \biggr]_{s=0}
\label{boundaryrel}
\end{equation}
where the integration is carried on all the variables $x^{\mu}$.
In turn, the relation \re{boundaryrel} implies conditions between the fields evaluated 
at $s=2\pi R$ and $s=0$. These boundary conditions will be written explicitly 
below for the case of a five dimensional warped space
without metric singularity or in Section \re{warpedspaceN} for the case with metric singularities.

We do not consider here the variable $s$ with a semi-infinite range $[0,\infty]$ 
(up to a transformation $s'=\pm\, s{+}\beta$) which require a special treatment. 

In order to obtain the boundary conditions which must be satisfied 
by the components $F^{[n,\Psi]}$ and $G^{[n,\Psi]}$ of the field 
(with identical relations for $F^{[n,\Phi]}$ and $G^{[n,\Phi]}$), one introduces 
the reduction ansatz \re{warp} in the boundary relation \re{boundaryrel}. 

As in the flat case for spinors \cite{GN4}, there are two sets of possible boundary conditions.
The first set
\begin{equation}
{\rm{Set\  BC1\ :}}
\hspace{0.5cm}\left\{\quad
\left(\matrix{F^{[n]}(2\pi R)\cr
              G^{[n]}(2\pi R)}\right)
              =B\left(\matrix{F^{[n]}(0)\cr
              G^{[n]}(0)}\right)
              \right.      
\label{caseBC1warp}
\end{equation}
where $B$ is a complex $2\times 2$ matrix. After some computation one finds that $B$ must be of the form
\begin{equation}
B=e^{4\epsilon\pi k R}\ e^{i\rho} \left(\matrix{\cosh(\omega)& \sinh(\omega)\cr
                                               \sinh(\omega)&\cosh(\omega)}\right)
\label{boundarymatrix}
\end{equation}
where $\rho$ is a real parameter with range $0\leq \rho <2\pi$ and $\omega$ is an arbitrary real parameter. 
Compared to the flat case there is simply an extra $e^{4\epsilon\pi k R}$ factor.

The second set
\begin{equation}
{\rm{Set\  BC2\ :}} \hspace{0.5cm}         
\left\{
\begin{array}{rcll}
G^{[n]}(0)&=&\epsilon_0 F^{[n]}(0)\quad\quad\ &,\quad  \epsilon_0^2=1
     \cr     
G^{[n]}(2\pi R)&=&\epsilon_R F^{[n]}(2\pi R) \quad&,\quad  \epsilon_R^2=1
\end{array}
\right.
\label{caseBC2warp}
\end{equation}
where $\epsilon_0$ and $\epsilon_R$ are two arbitrary signs is identical 
to the corresponding set in the flat case.

One  supposes that the fields satisfy the 
S0(3,1) invariant boundary conditions 

\subsection{Solutions. No metric singularity \label{solution}}

Introducing the reduction ansatz \re{warp} in the
five dimensional Dirac equation \re{Diracgen5} and postulating 
that $\psi^{[n]}$ satisfies 
the four dimensional parity invariant Dirac equation
\begin{equation}
(i\gamma^{\mu}\partial_{\mu}-m_n)\psi^{[n]}=0
\label{Dirac4}
\end{equation}
one finds from \re{dirac2warp} the two coupled equations for the components of the field
\bea
\partial_s G^{[n]}&=&\left(M-e^{\epsilon k s}\frac{m_n}{\lambda}\right)F^{[n]}+2\epsilon k G^{[n]}
     \nonumber\\
\partial_s F^{[n]}&=&2\epsilon k F^{[n]}+\left(M+e^{\epsilon k s}\frac{m_n}{\lambda}\right)G^{[n]}\ .
\label{warpeqfgs}
\eea
In terms of the following constants, variable and functions
\bea
{\widetilde{M}}&=&\frac{\epsilon M}{k}    
    \nonumber\\
{\widetilde{m}}_n&=&\frac{\epsilon m_n}{\lambda k}
    \nonumber\\
z&=&{\widetilde{m}}_n\,e^{\epsilon k s}
    \nonumber\\
{\widetilde{F}}^{[n]}(z)&=&F^{[n]}(s(z))
    \nonumber\\
{\widetilde{G}}^{[n]}(z)&=&G^{[n]}(s(z))
\label{changevareqfgz}
\eea
one obtains the simplified equations
\bea        
z\partial_z{\widetilde{F}}^{[n]}&=&2\,{\widetilde{F}}^{[n]}+\left({\widetilde{M}}+z\right)\,{\widetilde{G}}^{[n]}
        \nonumber\\
z\partial_z{\widetilde{G}}^{[n]}&=&2\,{\widetilde{G}}^{[n]}+\left({\widetilde{M}}-z\right)\,{\widetilde{F}}^{[n]}\ .
\label{newwarpeqfgz}
\eea
Defining
\bea 
{\widetilde{F}}_+^{[n]} &=& {\widetilde{F}}^{[n]}+{\widetilde{G}}^{[n]}
     \nonumber\\
{\widetilde{F}}_-^{[n]} &=& {\widetilde{F}}^{[n]}-{\widetilde{G}}^{[n]}\ ,
\label{fplusdef}
\eea 
one equation gives ${\widetilde{F}}_-^{[n]}$ in terms of ${\widetilde{F}}_+^{[n]}$ 
\begin{equation}
 {\widetilde{F}}_-^{[n]} =
z^{\frac{3}{2}}
\left( -  z \partial_z {\widetilde{F}}_+^{[n]} + \left({\widetilde{M}} -\frac{1}{2}\right){\widetilde{F}}_+^{[n]} \right)
\label{fmoinsplus} 
\end{equation}
while the second equation, of second order, leads to
\begin{equation}
z^2\partial^2_z {\widetilde{F}}_+^{[n]} - 4 z\partial_z {\widetilde{F}}_+^{[n]} 
+ \left(z^2+6- {\widetilde{M}}^2 + {\widetilde{M}}\right){\widetilde{F}}_+^{[n]}=0\ .
\label{newwarpplus1}
\end{equation}
The function ${\cal{F}}_+^{[n]}=z^{-\frac{5}{2}}{\widetilde{F}}_+^{[n]}$ satisfies the following equation
\begin{equation}
z^2\partial^2_z {\cal{F}}_+^{[n]} + z\partial_z {\cal{F}}_+^{[n]} 
+ \left(z^2-({\widetilde{M}}-{\frac{1}{2}})^2\right){\cal{F}}_+^{[n]}=0
\label{newwarpplus2}
\end{equation}
which is of Bessel type. The final solution for $F^{[n]}_+(s)$ is the following linear superposition
of Bessel functions.
\begin{equation}
F^{[n]}_+(s)=\left(\frac{ m_n e^{\epsilon k s}}{\lambda k}\right)^{\frac{5}{2}}
     \left(
     \sigma_n J_{\frac{\epsilon M}{k}-\frac{1}{2}}\left( \frac{ m_n e^{\epsilon k s}}{\lambda k}\right)    
   + \tau_n Y_{\frac{\epsilon M}{k}-\frac{1}{2}}\left( \frac{ m_n e^{\epsilon k s}}{\lambda k}\right) 
     \right)
\label{solnewwarplus}
\end{equation}
with two arbitrary constants while
\begin{equation}
 F_-^{[n]}(s) =
\frac{\lambda e^{-\epsilon k s}}{m_n}
\left( - \partial_s   F_+^{[n]}(s) 
+ \left( M  +2\epsilon k \right)F_+^{[n]}(s) \right)
\label{fmoinsplusb} 
\end{equation}   
and
\bea
 F^{[n]}(s)&=&\frac{1}{2}\left(F^{[n]}_+(s)+F^{[n]}_-(s)\right)    
     \nonumber\\
 G^{[n]}(s)&=&\frac{1}{2}\left(F^{[n]}_+(s)-F^{[n]}_-(s)\right)\ .         
\label{finalfg}
\eea

The constants $\sigma_n$ and $\tau_n$
of \re{solnewwarplus} 
are determined by the boundary conditions \re{caseBC1warp} or \re{caseBC2warp}.

\newpage
\section{Application to a five dimensional warped space
with metric singularities \label{warpedspaceN}}

\subsection{The five-dimensional warped space
with $N$ metric singularities  \label{FivedimN}}

The extension of the preceding arguments
to a warped space 
with an arbitrary number $N$ of metric singularities situated at the points
$s_i,\,i=1,N$ with
$s_0=0<s_1<s_2,\dots,s_N<s_{N+1}=2\pi R$ on the strip is straightforward. 
By definition, the metric is of the general form \re{ds2warp}
with $\epsilon=1$ for some (possibly non connected) region of $s$ and $\epsilon=-1$ for the complementary region.
A singular point is a point which joins two regions of opposite values of
$\epsilon$.
We moreover postulate,
for physical reasons, that all the components of the metric 
are continuous at the singular points. 

There are $N+1$ intervals $I_i,\ i=0,\dots,N$ 
\begin{equation} 
I_0=[0,s_1],\ I_1=[s_1,s_2],\ \dots\ ,\ I_{N-1}
=[s_{N{-}1},s_N],\ I_N=[s_N,2\pi R] 
\label{intervals}
\end{equation}
of respective length
\begin{equation}
l_0=s_1,\,l_1=s_2{-}s_1,\,l_2=s_3{-}s_2,\,\dots,\,l_N=2\pi R{-}s_N \ .
\label{lengths}
\end{equation}
Defining
\begin{equation} 
r_i=-2(-1)^{i{+}1}\left(\sum_{j=0}^{i-1}(-1)^j s_{i-j}\right)
    \label{theri1}
\end{equation}
(note $r_0 {=} 0$) equivalent to    
\bea
r_{2i}&=&2\sum_{j=1}^{i} l_{2j{-}1}
    \nonumber\\
r_{2i+1}&=&-2\sum_{j=0}^{i} l_{2j}  \ ,
\label{theri2}
\eea
the metric takes the explicit form  
\begin{equation}
{\rm{for\ }}s\in I_i \quad :\quad dS^2
=e^{-2k\epsilon \left((-1)^{i}s-r_i\right)} dx_{\mu}dx^{\mu}-ds^2\quad (i=0,\dots,N)
\label{manysing}     
\end{equation}
chosen to be normalised to one at $s=0$.
The sign of the coefficient of $s$ in the exponent alternates 
between $\epsilon$ and $-\epsilon$ for the intervals $I_i$ 
with even and odd $i$. 
The end points of each interval 
are thus singular points, except $s=0$ and $s=2\pi R$ 
(see however the special case of a closure to a circle in Sec.\re{circle}).

The solution for the non zero elements of the vielbein, of the vector connection, 
of the covariant derivatives of the vielbein and of the spinor connection 
depends on the intervals. For $s\in I_i$ 
\be
\begin{array}{rcl}
e_{\Sigma}^{\phantom{D}\sigma}
=e_{\displaystyle{0}}^{\phantom{D}0}=e_{\displaystyle{1}}^{\phantom{D}1}
=e_{\displaystyle{2}}^{\phantom{D}2}=e_{\displaystyle{3}}^{\phantom{D}3}
&=& e^{-\epsilon k((-1)^is-r_i)}
\quad,\quad e_{S}^{\phantom{D}s}=1
        \vspace{0.2 cm}
        \cr
e_{\sigma}^{\phantom{D}\Sigma}
=e_{0}^{\phantom{D}\displaystyle{0}}=e_{1}^{\phantom{D}\displaystyle{1}}
=e_{2}^{\phantom{D}\displaystyle{2}}=e_{3}^{\phantom{D}\displaystyle{3}}
&=&e^{\epsilon k((-1)^is-r_i)}
\quad,\quad e_{s}^{\phantom{D}S}=1
        \vspace{0.2 cm}
        \cr
G_{\displaystyle{500}}
     =G_{\displaystyle{050}}
     =-G_{\displaystyle{005}}
     =-G_{\displaystyle{5II}}&
                          \cr
     =-G_{\displaystyle{I5I}}
     =G_{\displaystyle{II5}}
     &=&(-1)^i\epsilon ke^{-2\epsilon k((-1)^is-r_i)}       
        \vspace{0.2 cm}
       \cr
G_{\displaystyle{50}}^{\phantom{AB}\displaystyle{0}}
    =G_{\displaystyle{05}}^{\phantom{AB}\displaystyle{0}}
    =G_{\displaystyle{5I}}^{\phantom{AB}\displaystyle{I}}
    =G_{\displaystyle{I5}}^{\phantom{AB}\displaystyle{I}}
    &=&(-1)^i\epsilon k 
        \vspace{0.2 cm}
       \cr
G_{\displaystyle{00}}^{\phantom{AB}\displaystyle{5}}
    =-G_{\displaystyle{II}}^{\phantom{AB}\displaystyle{5}}
    &=&(-1)^i\epsilon ke^{-2\epsilon k((-1)^is-r_i)} 
        \vspace{0.2 cm}
       \cr
\left(D_{\displaystyle{0}}e\right)_{\displaystyle{5}}^{\phantom{D}o}
= \left(D_{\displaystyle{I}}e\right)_{\displaystyle{5}}^{\phantom{D}i}
&=&(-1)^i\epsilon ke^{-\epsilon k((-1)^is-r_i)}
        \vspace{0.2 cm}
        \cr
\left(D_{\displaystyle{0}}e\right)_{\displaystyle{0}}^{\phantom{D}5}
= -\left(D_{\displaystyle{I}}e\right)_{\displaystyle{I}}^{\phantom{D}5}
&=&(-1)^i\epsilon ke^{-2\epsilon k((-1)^is-r_i)}    
        \vspace{0.2 cm}
         \cr
G^{[s]}_{\Sigma}&=&
-\frac{1}{2}(-1)^i\epsilon ke^{-\epsilon k((-1)^is-r_i)} \gamma_{\sigma}\gamma^5 \ .
\end{array}
\label{vielbeinN}
\ee

In each subspace, the Dirac equation derived from \re{Diracgen5} assumes the form
\be
{\rm{for\ }}s\in I_i\quad:\quad
\biggl(e^{\epsilon k((-1)^is-r_i)}\left(i\gamma^{\sigma}\partial_{\Sigma}\right)
+(i\gamma^5)\left(\partial_{\displaystyle{5}}-2(-1)^i k\epsilon\right)\biggr)\Psi=M\Psi \ .
\label{DiracwarpN}
\ee

\subsection{Kaluza-Klein reduction
with metric singularities  \label{KKreducN}}

We adopt the following Kaluza-Klein separation of variables
analogous to the no singularity case \re{warp}
\bea
\Psi(x^{\mu},s)
&=&{\displaystyle{\sum_n}} \psi^{[n]}(x^{\mu},s)
    \nonumber\\
&=&{\displaystyle{\sum_n}} \biggl(F^{[n]}(s)+i\,G^{[n]}(s)\gamma^{5}\biggr)\psi^{[n]}(x^{\mu})
\label{warpN}
\eea
assuming 
$\psi^{[n]}(x^{\mu})$ to be a spinor depending on $x^{\mu}$ only, and independent of the interval $I_i$
to which $s$ belongs.
The complex functions $F^{[n]}(s)$ and $G^{[n]}(s)$ are functions depending on $s$ only.
They are supposed to be smooth within the intervals $I_i$,  
where they may take different analytical forms, respectively $F^{[n,i]}(s)$ and $G^{[n,i]}(s)$.

Introducing the reduction \re{warpN}
into the Dirac equation \re{DiracwarpN} in each subspace $I_i$ and 
postulating that the $\psi^{[n]}(x_{\mu})$ satisfies the four dimensional Dirac equation \re{Dirac4}, 
we find the two coupled equations  
\begin{equation}
{\rm{for\ }}s\in I_i\ 
\left\{
\begin{array}{rcl}
\partial_s G^{[n,i]}&=&\left(M-e^{\epsilon k((-1)^is-r_i)}m_n\right)F^{[n,i]}+2(-1)^i\epsilon k G^{[n,i]}
     \cr
\partial_s F^{[n,i]}&=&2(-1)^i\epsilon k F^{[n,i]}+\left(M+e^{\epsilon k((-1)^is-r_i)}m_n\right)G^{[n,i]} \ .
\end{array}
\right.
\label{warpNeqfgs}
\end{equation}

\subsection{Solutions with metric singularities \label{solutiuonN}}

Following the same procedure as in the case without singularity 
\re{solution} we find
that in the interval $I_i$, the solution is
\bea
 F^{[n,i]}(s)&=&\frac{1}{2}\left(F^{[n,i]}_+(s)+F^{[n,i]}_-(s)\right)    
     \nonumber\\
 G^{[n,i]}(s)&=&\frac{1}{2}\left(F^{[n,i]}_+(s)-F^{[n,i]}_-(s)\right)\ .         
\label{finalfgi}
\eea
The function $F^{[n,i]}_+(s)$ is the following linear superposition
of Bessel functions.
\bea
F^{[n,i]}_+(s)&=&\left(\frac{ m_n  e^{ \epsilon k\left((-1)^i s-r_i\right)}}{k}\right)^{\frac{5}{2}}
     \Biggl(
     \sigma_{n,i}\  
     J_{\frac{\epsilon (-1)^i M}{k}-\frac{1}{2}}\!\!
     \left({\scriptstyle{ \frac{ m_n e^{\epsilon k((-1)^i s-r_i)}}{ k}}}\right)\Biggr.    
    \nonumber\\
   &&\Biggl.+ \tau_{n,i}\  
   Y_{\frac{\epsilon (-1)^i M}{k}-\frac{1}{2}}\!\!
   \left({\scriptstyle{ \frac{ m_n  e^{\epsilon k ((-1)^i s-r_i)}}{ k}}}\right) 
     \Biggr)
\label{solnewwarplusi}
\eea
depending on two arbitrary constants $\sigma_{n,i},\tau_{n,i}$ and with
\begin{equation}
 F_-^{[n,i]}(s) =
\frac{ e^{-\epsilon k \left((-1)^is-r_i\right)}}{m_n}
\left( - \partial_s   F_+^{[n,i]}(s) 
+ \Biggl( M  +2\epsilon k (-1)^i\Biggr)F_+^{[n,i]}(s) \right)\ .
\label{fmoinsplusbi} 
\end{equation}   

The constants $\sigma_{n,i}$ and $\tau_{n,i}$ (altogether $2(N+1)$ parameters)
must satisfy 
$2(N+1)$ homogeneous linear boundary relations 
expressing the boundary conditions (see \re{BCPred}).
For given boundary conditions, in order to obtain a non trivial solution for the 
$\sigma_{n,i}$ and $\tau_{n,i}$, the related $2(N{+}1)\times 2(N{+}1)$ determinant
must vanish, leading to a mass eigenvalue equation for the $m_n$.

\subsection{Boundary relation and conditions for the spinor fields 
with metric singularities \label{boundrelN}}

In the boundary relation \re{integral}, the total derivative terms in $\Sigma$ vanish
since the fields are supposed to decrease sufficiently fast at infinity in the $\Sigma$ 
directions. The fields are in general discontinuous at the metric singularity point. 
We define
\bea
\Psi^l(s_i)&=&\lim_{\eta\rightarrow 0^+}\Psi(x^{\mu},s_i-\eta)
    \nonumber\\
\Psi^r(s_i)&=&\lim_{\eta\rightarrow 0^+}\Psi(x^{\mu},s_i+\eta)\ .
\label{psirl}
\eea
Noting that $\gamma^{\displaystyle{5}}=\gamma^5$ \re{gammawarp} and \re{vielbeinN},
the boundary relation \re{integral}, 
after integration over $s$, becomes 
\be
\int d^4x\Biggl(
\sum_{i=1}^{N+1} \overline\Phi^l(s_i)\gamma^5\Psi^l(s_i)\sqrt{g(s_i)}
-\sum_{i=0}^{N} \overline\Phi^r(s_i)\gamma^5\Psi^r(s_i)\sqrt{g(s_i)}
    \Biggr)=0 \ .
\label{BCrelN}    
\ee
Expanding $\Psi$ and $\Phi$ 
according to the Kaluza-Klein reduction \re{warpN}, leading to
\begin{equation}
{\overline{\Phi}}(x^{\mu},s)=\sum_n {\overline{\phi}}^{[n]}(x^{\mu})\biggl(C^{*[n]}(s)-iD^{*[n]}(s)\gamma^{5}\biggr)
\ ,
\label{warpphi}
\end{equation}
one finds after some algebra
\bea
\sum_{i=0}^{N} \biggl(D^{[m]*r}(s_i)F^{[n]r}(s_i)-C^{[m]*r}(s_i)G^{[n]r}(s_i)
                 \biggr)\sqrt{g(s_i)}\hspace*{1 cm}&&
                 \nonumber\\
-\sum_{i=1}^{N+1} \biggl(D^{[m]*l}(s_i)F^{[n]l}(s_i)-C^{[m]*l}(s_i)G^{[n]l}(s_i)
                 \biggr)\sqrt{g(s_i)}&=&0
     \label{relN1}\\
\sum_{i=0}^{N} \biggl(C^{[m]*r}(s_i)F^{[n]r}(s_i)-D^{[m]*r}(s_i)G^{[n]r}(s_i)
                 \biggr)\sqrt{g(s_i)}\hspace*{1 cm}&&
                 \nonumber\\
-\sum_{i=1}^{N+1} \biggl(C^{[m]*l}(s_i)F^{[n]l}(s_i)-D^{[m]*l}(s_i)G^{[n]l}(s_i)
                 \biggr)\sqrt{g(s_i)}&=&0\ .
\label{relN2}
\eea

In terms of the left and right boundary values, we define the $4(N{+}1)$ dimensional vectors
\be
\Phi=\left(
      \begin{array}{c}
      {\scriptstyle{\sqrt[4]{g(s_0)}}}\,C^{[n]r}(s_0)\cr
      {\scriptstyle{\sqrt[4]{g(s_0)}}}\,D^{[n]r}(s_0)\cr
      {\scriptstyle{\sqrt[4]{g(s_1)}}}\,C^{[n]r}(s_1)\cr
      {\scriptstyle{\sqrt[4]{g(s_1)}}}\,D^{[n]r}(s_1)\cr
      \vdots\cr
      {\scriptstyle{\sqrt[4]{g(s_N)}}}\,C^{[n]r}(s_N)\cr      
      {\scriptstyle{\sqrt[4]{g(s_N)}}}\,D^{[n]r}(s_N)\cr      
      {\scriptstyle{\sqrt[4]{g(s_1)}}}\,C^{[n]l}(s_1)\cr
      {\scriptstyle{\sqrt[4]{g(s_1)}}}\,D^{[n]l}(s_1)\cr
      \vdots\cr
      {\scriptstyle{\sqrt[4]{g(s_N)}}}\,C^{[n]l}(s_N)\cr
      {\scriptstyle{\sqrt[4]{g(s_N)}}}\,D^{[n]l}(s_N)\cr
      {\scriptstyle{\sqrt[4]{g(s_{N+1})}}}\,C^{[n]l}(s_{N+1})\cr
      {\scriptstyle{\sqrt[4]{g(s_{N+1})}}}\,D^{[n]l}(s_{N+1})
           \end{array}
     \right)
     \quad,\quad
\Psi=\left(\begin{array}{c}
      {\scriptstyle{\sqrt[4]{g(s_0)}}}\,F^{[n]r}(s_0)\cr
      {\scriptstyle{\sqrt[4]{g(s_0)}}}\,G^{[n]r}(s_0)\cr
      {\scriptstyle{\sqrt[4]{g(s_1)}}}\,F^{[n]r}(s_1)\cr
      {\scriptstyle{\sqrt[4]{g(s_1)}}}\,G^{[n]r}(s_1)\cr
      \vdots\cr
      {\scriptstyle{\sqrt[4]{g(s_N)}}}\,F^{[n]r}(s_N)\cr      
      {\scriptstyle{\sqrt[4]{g(s_N)}}}\,G^{[n]r}(s_N)\cr      
      {\scriptstyle{\sqrt[4]{g(s_1)}}}\,F^{[n]l}(s_1)\cr
      {\scriptstyle{\sqrt[4]{g(s_1)}}}\,G^{[n]l}(s_1)\cr
      \vdots\cr
      {\scriptstyle{\sqrt[4]{g(s_N)}}}\,F^{[n]l}(s_N)\cr
      {\scriptstyle{\sqrt[4]{g(s_N)}}}\,G^{[n]l}(s_N)\cr
      {\scriptstyle{\sqrt[4]{g(s_{N+1})}}}\,F^{[n]l}(s_{N+1})\cr
      {\scriptstyle{\sqrt[4]{g(s_{N+1})}}}\,G^{[n]l}(s_{N+1})
      \end{array}
      \right) \ .
\label{matrixpsi}
\ee
The two boundary relations \re{relN1}, \re{relN2} can be written in matrix form
\be
\Phi^{+}S^{[4(N{+}1)]}_j\Psi=0 \quad,\quad j=1,2
\label{relmat}
\ee
where $S_j$ are square matrices 
with upper index $[4(N{+}1)]$ referring to their size. 
For \re{relN1}, the antisymmetric matrix $S^{[4(N+1)]}_1$ has the following form 
\be
S^{[4(N+1)]}_1=\left(
    \matrix{  S^{[2(N+1)]}_1    &  \phantom{-}0^{[2(N+1)]} \cr
             0^{[2(N+1)]}       & -S^{[2(N+1)]}_1
           }
    \right)
\label{matrS21a}
\ee    
with the zero matrix $0^{[2(N+1)]}$ and
the antisymmetric block diagonal matrix $S^{[2(N+1)]}_1$ 
\be
S^{[2(N+1)]}_1=\left(
   \matrix{{-i\sigma_2}& 0^{[2]}             &\dots    \cr
         0^{[2]}           & {-i\sigma_2}  &\dots    \cr
         \vdots            &\vdots               &\ddots
       }
               \right) \ .
\label{matrS11}
\ee
For \re{relN2}, the matrix $S^{[4(N+1)]}_2$ is block diagonal 
\be
S^{[4(N+1)]}_2=\left(
    \matrix{  S^{[2(N+1)]}_2           &  \phantom{-}0^{[2(N+1)]} \cr
             0^{[2(N+1)]}   & -S^{[2(N+1)]}_2
           }
    \right)
\label{matrS21b}
\ee    
with the diagonal matrix $S^{[2(N+1)]}_2$
\be
S^{[2(N+1)]}_2=\left(
   \matrix{\sigma_3     & 0^{[2]}   &\dots    \cr
           0^{[2]}      & \sigma_3  &\dots    \cr
           \vdots       &\vdots     &\ddots
       }
               \right) \ .
\label{matrS21c}
\ee

The allowed sets of boundary conditions can be obtained 
from the boundary relations \re{relmat}, 
by the following general procedure. The boundary conditions are 
expressible in terms of $2(N{+}2)$
independent homogeneous linear relations among the components of the matrix $\Psi$ \re{matrixpsi}
chosen in such a way as to guarantee the two boundary relations \re{relN1} and \re{relN2}.
The components of $\Phi$ have of course to satisfy the same linear relations. 
The boundary conditions are written 
\be
M_{BC}\,\Psi=0
\label{BCmatM}
\ee
where $M_{BC}$ is a $2(N{+}1)\times 4(N{+}1)$ matrix of rank $2(N{+}1)$. 
For any such $M_{BC}$, there exists a $4(N{+}1)\times 4(N{+}1)$ permutation matrix $P$ such that,
defining 
\be
\Psi_P\equiv P\,\Psi        \ ,
\label{PsiP}
\ee
the $2(N{+}1)$ boundary conditions are equivalent to
\be
\Psi_P=V^{[4(N{+}1)]}_P \, \Psi_P 
\label{BCmatV}
\ee
with the matrix $V^{[4(N{+}1)]}_P$ written in terms of 
a matrix $V^{[2(N{+}1)]}_P$ (depending on $P$) and the unit matrix
$1^{[2(N+1)]}$ as
\be
V^{[4(N{+}1)]}_P =\left(
   \matrix{1^{[2(N{+}1)]}   & 0^{[2(N{+}1)]}\cr  
           V^{[2(N{+}1)]}_P & 0^{[2(N{+}1)]}        
          }       
                  \right) \ .    
\label{PsiV}
\ee
 
Writing $\Psi_P$ in terms of its $2(N{+}1)$ upper elements $\Psi_P^u$ and its
down elements $\Psi_P^d$
\be
\Psi_P=\left(
       \matrix {\Psi_P^u\cr
                \Psi_P^d
                }
       \right)
\label{vectorpsiP}
\ee
one finds that the $2(N{+}1)$ first equations are trivial while the 
last $2(N{+}1)$ equations express the boundary conditions 
equivalent to \re{BCmatM}
\be
\Psi_P^d=V^{[2(N{+}1)]}_P\,\Psi_P^u\ .
\label{BCPred}
\ee
This is in agreement with the observation that, 
from \re{BCmatM}, there exists 
always a permutation $P$ of the component of $\Psi$ such that 
the $2(N{+}1)$ components ($\Psi_P^d$) are linear functions
of the $2(N{+}1)$ other components ($\Psi_P^u$).

Writing $S^{[4(N+1)]}_{Pj}$ ($j=1,2$)the transformed of $S^{[4(N+1)]}_j$ under $P$
\be
S^{[4(N+1)]}_{Pj}=P\,S^{[4(N+1)]}_{j}\,P^{-1}\ ,
\label{SjP}
\ee
the matrix $V^{[4(N+1)]}_{P}$ expressing the allowed boundary conditions
\re{BCmatV},
must satisfy the two matrix equations
\be
\left(V^{[4(N+1)]}_{P}\right)^+\, S^{[4(N+1)]}_{Pj}\, V^{[4(N+1)]}_{P}=0 \ .
\label{BCPj}
\ee
This follows from the fact that the boundary relations \re{relmat} 
then depend on $\Psi^u_P$ and $\Phi^{u+}_P$ only 
which are all independent and arbitrary.

With the four $2(N+1)\times 2(N+1)$ matrices 
$S_{Pj}^{[2(N+1)],r},\, r=1,\dots,4,\,j=1,2$ 
defined for each $S_{Pj}^{[4(N+1)]}$ as
\be
S_{Pj}^{[4(N+1)]}=\left(
         \matrix{S_{Pj}^{[2(N+1)],1}  &  S_{Pj}^{[2(N+1)],2}   \cr
                 S_{Pj}^{[2(N+1)],3}  &  S_{Pj}^{[2(N+1)],4}
                }
         \right) \ ,
\label{SPjr}
\ee
the boundary relations \re{relmat}
lead explicitly to two equations for 
the matrix $V_{P}^{[2(N+1)]}$
\bea
&{\rm{for\ }}j=1,2\phantom{\Bigl[\Bigr.}&         
         \nonumber\\
&S_{Pj}^{[2(N+1)],1}+\left(V^{[2(N+1)]}_{P})\right)^+\, S_{Pj}^{[2(N+1)],3}
+S_{Pj}^{[2(N+1)],2}\,V^{[2(N+1)]}_{P}&
         \nonumber\\
&\hspace{3 cm}+\left(V^{[2(N+1)]}_{P}\right)^+\,S_{Pj}^{[2(N+1)],4}\,V^{[2(N+1)]}_{P}=0 \ .&
\label{relVP}
\eea

It should be stressed that different choices of $P$ may lead to equivalent, 
differently expressed,  boundary conditions, in particular by 
multiplying $P$ by further permutations within the elements of $\Psi^u$ or within the elements of $\Psi^d$.
A few examples of boundary conditions are given in App.\re{ExempBC}.

\section{Physical considerations \label{physdis1}}

In our previous article \cite{GN3}, we have given a detailed discussion
of the physical relevance of the main aspects 
underlying the Kaluza-Klein construction for scalars. 
We summarize here the points which apply to the spinor case. 

\subsection{Physical discussion of the generalized
boundary conditions \label{locality}}

It happens that the boundary conditions \re{BCPred} may connect the values of the 
components $F$ and $G$ of the field 
(not their derivatives as in the scalar case) 
at different points of the $s$-domain
i.e. at the $N$ metric singular points and at the two edges. In this case, the field
explores, in fact, its full domain at once. This is tantamount to action at a distance
or non locality, which we argued in \cite{GN3} not
to be in contradiction with quantum mechanics.

In our numerical applications \re{Numeric} however, we restrict ourselves 
either to fully local boundary conditions 
(locality at the metric singular points as well as at the edges (App.\ref{localBC}))
or to partially local boundary conditions (excluding locality at the edges (App.\ref{semilocalBC})). 

\subsection{Closure to a circle \label{circle}}

The strip could be closed in a circle by identifying the points $s=0$ and $s=2\pi R$ 
under the following requirements.

There must be an even number $2p\ (p>0)$ of singularities. By rotation around the circle,
the first singularity can always be 
placed at the closure point $s=0$. Then $N\equiv 2p{-}1$.
The total range where the sign of $s$ in the exponential in the metric 
is positive must be equal to the total range where it is negative \re{intervals},
\re{lengths}
\begin{equation}
\sum_{j=0}^{j=p-1}l_{2j}=\sum_{j=0}^{j=p-1}l_{2j+1}=\pi R \ .
\label{closurecond}
\end{equation}

\subsection{The ''one-mass-scale-only'' hypothesis \label{masscale}}

By assumption,
there is only one high mass scale in the theory
which is chosen to be the Planck mass
\begin{equation}
M_{{\rm{Pl}}}\approx 1.22\ 10^{16}\ {\rm{TeV}}\ .
\label{Planck}
\end{equation}
The dimensionfull parameters $k$, $R$ and $M$ can be written in terms of reduced parameters
${\overline{k}}$, ${\overline{R}}$ and ${\overline{M}}$
\bea
k&=&{\overline{k}}\,M_{{\rm{Pl}}}      
\nonumber\\
R&=&{\overline{R}}\,\left(M_{{\rm{Pl}}}\right)^{-1}
\nonumber\\
M&=&{\overline{M}}\,M_{{\rm{Pl}}}   \ .    
\label{barpara}
\eea
We call the assumption that the reduced parameters 
are neither
large nor small numbers (except 0) the 
``one-mass-scale-only'' hypothesis. 
The parameter
$\overline{k}\overline{R}\,=\,kR$ governs the reduction from the high mass scale to the 
TeV scale of the low lying masses in the Kaluza-Klein towers.

Finally, let us note that by rescaling the parameter 
${\overline{k}}$ can always be chosen to be equal to one
\begin{equation}
{\overline{k}}=1\ .
\label{kequalone} 
\end{equation}
Since the mass eigenvalue equation are covariant under a rescaling
of all the reduced parameters ${\overline{p}}$ according to their energy dimension $d_p$
\begin{equation}
{\overline{p}}\rightarrow \lambda^d_p{\overline{p}}\ ,
\label{rescaling1}
\end{equation}
one finds that the mass eigenvalues for a given ${\overline{k}}$ can be obtained from
eigenvalues corresponding to our choice ${\overline{k}}=1$ (using 
$\lambda=1/{\overline{k}}$) by
\begin{equation}
m_n\Biggl(\Biggl\{{\overline{k}}, {\overline{R}},{\overline{M}}  \Biggr\}\Biggr)
={\overline{k}}m_n\Biggl(\Biggl\{1,{\overline{k}} {\overline{R}},\frac{\overline{M}}{\overline{k}}  \Biggr\}\Biggr) \ .
\label{rescaling2}
\end{equation}

\subsection{The physical masses \label{massphys}}

For a four-dimensional observer supposed to be sitting at $s=s_{\rm{obs}}$
in a given $I_i$ interval \re{intervals},
the metric \re{manysing} 
\begin{equation}
dS^2=e^{-2\epsilon k((-1)^i s_{\rm{obs}}-r_i)}dx_{\mu} dx^{\mu}-ds^2
\label{canonicmetric}
\end{equation}
can be transformed 
in canonical form
\begin{equation}
dS^2=d{\widetilde{x}}_{\mu} d{\widetilde{x}}^{\mu}-ds^2
\label{canonicmetric2}
\end{equation}
by the following rescaling 
\begin{equation}
{\widetilde{x}}_{\mu}=e^{-\epsilon k((-1)^i s_{\rm{obs}}-r_i)}x_{\mu}\ .
\label{canonictf}
\end{equation}
According to \re{Dirac4}, we have
\begin{equation}
i\gamma^{\mu}{\widetilde{\partial_{\mu}}}\psi^{[n]}
=e^{\epsilon k((-1)^i s_{\rm{obs}}-r_i)}
  (i\gamma^{\mu}\partial_{\mu}\psi^{[n]})
  =e^{\epsilon k((-1)^i s_{\rm{obs}}-r_i)}m_n\psi^{[n]} \ .
\label{Dirac4obs}
\end{equation}
The mass as seen in by the observer in the brane at $s=s_{\rm{obs}}\in I_i$ is 
thus related to the mass eigenvalue $m_n$ by
\begin{equation}
m_n^{{\rm{obs}}}=e^{\epsilon k\left((-1)^i s_{{\rm{obs}}}-r_i\right)}\,m_n\ .
\label{massquare2}
\end{equation}
For $s_{\rm{obs}}=0$, the physical mass 
is just equal to the mass eigenvalue. 

\subsection{Probability density \label{densprob}}

Once all the parameters defining a specific model are chosen and the
mass eigenvalue tower is determined, there exists
a unique field $\psi^{[n]}(x^{\mu},s)$ (see \re{warpN})
for each mass eigenvalue
leading to a naive
probability density field distribution $D^{[n]}(x^{\mu},s)$
which depends both on $x^{\mu}$ and $s$
\begin{equation}
D^{[n]}(x^{\mu},s)=\sqrt{g}\,\left(\overline{\psi}^{[n]}(x^{\mu},s)\psi^{[n]}(x^{\mu},s)\right)\ .
\label{probdensity}
\end{equation}
Note that the shape of this density distribution
depends in general on the interval $I_i$
to which $s$ belongs.
As observed and discussed in \cite{GN2}, 
these probability densities are fast varying functions of $s$.
The total normalized probability density 
for a Kaluza-Klein particle to be present 
in a brane situated at $s=s_{\rm{obs}}$ is
\begin{equation}
D^{[n]}(s_{\rm{obs}})=\frac{\int d^4x\, D^{[n]}(x^{\mu},s_{\rm{obs}})}
                       {\int d^4x\, ds\, D^{[n]}(x^{\mu},s)}\ .
\label{probdensobs}
\end{equation}
Remember that the physical mass as seen by the observer 
is also a function of the $s_{\rm{obs}}$ position \re{massquare2}. 

\section{Towers \label{towers}}

In the absence of metric singularities, the two arbitrary parameters 
$\sigma$ and $\tau$ which appear in the solution \re{finalfg}, \re{solnewwarplus}, \re{fmoinsplusb} 
of the Dirac equation \re{Diracgen5} in the five dimensional space
after the KK reduction \re{warp} have to satisfy two
homogeneous linear equations expressing an allowed set of boundary conditions,
belonging either to the set BC1 \re{caseBC1warp} or to the set BC2 \re{caseBC2warp}.
The condition for the existence of a non trivial $\sigma$, $\tau$ solution is
the vanishing of the related determinant. This leads 
in each case to a mass 
equations from which the KK mass towers can be derived. In Sect.\re{nosingtower}, numerical 
examples of KK mass towers are given for each of the two sets of boundary conditions, for different 
values of the basic parameters of the model, i.e. the warp factors $\epsilon,\,k,\,kR$, the bulk 
mass $M$, as well as for different values of the parameters 
$\rho, \omega$ or $\epsilon_0,\, \epsilon_R$ defining the boundary conditions 
considered.

In the general case, when there are $N$ metric singularities,
there are $N{+}1$ parameters $\sigma_{n,i}$ and $N{+}1$ parameters $\tau_{n,i}$
appearing
in the solution \re{finalfg}, \re{solnewwarplusi}, \re{fmoinsplusbi}
of the Dirac equation after the Kaluza-Klein reduction
\re{warpN}.
These parameters have to satisfy the $2(N{+}1)$ homogeneous 
linear equations \re{BCPred} resulting from the imposition of the 
$2(N{+}1)$ boundary conditions on the $2(N{+}1)$ values of the fields 
at the edges of the $N+1$
intervals $I_i$ in the $s$-range \re{psirl}. 
Indeed, for a 
given singularity 
configuration, there exists a set of  $2(N+1)$ boundary conditions resulting from the two boundary 
relations \re{relmat} expressing the condition of hermiticity of the Dirac operator.
As in the preceding case, the requested vanishing of the determinant of the coefficients
of the  $2(N+1)$ boundary conditions with respect to the $2(N+1)$ 
parameters ($\sigma_{n,i}$, $\tau_{n,i}$)
leads to the corresponding KK mass equation. In Sect.\re{singtower}, 
a few examples of towers are given when there is one singularity. 

For completeness, let us list all the parameters. They are the basic parameters of the 
warp model
$k,\epsilon, kR$, the bulk mass $M$, the positions $s_i$ of the $N$ metric singularities and the 
boundary parameters defining the
matrix $V_P^{[2(N+1)]}$ subject to the two conditions \re{relVP}. 
Once all these parameters are chosen, the vanishing of the above determinant is 
generally a transcendental function of
the eigenvalues $m_n$. 
  
\section{Examples of towers \label{Numeric}}

For an illustration of the types of spinor towers which appear in warped spaces, we construct examples
of the eight lowest mass eigenvalues for simple specific boundary conditions. 
We first discuss the case when there is no metric singularity, then when there is one metric singularity.
We would like to stress that, in order to perform the numerical computations, high precision is mandatory.

\subsection{Examples of towers. No metric singularity  \label{nosingtower}}

In this subsection, a few illustrative numerical examples of Kaluza-Klein spinor towers 
in warped spaces are presented for each of the two sets BC1 \re{caseBC1warp} 
and  BC2 \re{caseBC2warp} of boundary conditions and for some chosen values 
of the bulk mass $M$ and of the parameters fixing the boundary conditions. 
In general, the Kaluza-Klein mass eigenvalues are irregularly spaced. 
With the adopted values of the basic parameters of the warp model, 
i.e. $k$ arbitrarily normalized to the Planck mass ($\overline{k}=1$, see \re{barpara})
and $kR\approx 6.3$, all the low lying 
Kaluza-Klein masses are of the order TeV. 
In the tables, $kr$ is fixed to
\begin{equation}
kR=6.3
\label{kR}
\end{equation}
and the Kaluza-Klein tower masses denoted with $\widetilde{m}_i$ are given in TeV. 
As a general rule, the values of  $\widetilde{m}_i$ decrease (exponentially)
when $kR$ increases, hence fixing the overall scale of the masses in the tower. 
It should be noted that chosing the value of the bulk mass $M$ 
to zero or to values of the order of the Planck mass, 
within the one-mass-scale-only \re{masscale}, does not 
lead to substantially different Kaluza-Klein towers.

In Table\re{tableM1}, the eight low lying mass eigenvalues
($\widetilde{m}_i,\ i=1,\dots,8$) 
of Kaluza-Klein towers 
are given in the case of boundary conditions BC1 \re{caseBC1warp} 
for zero bulk mass $M$, for different values of the parameter $\rho$, 
and for each of them, 
for different values of the parameter $\omega$ \re{boundarymatrix}.
One observe that the first mass of the towers, $\widetilde{m}_1$, 
is relatively sensitive to the value of $\omega$, 
particularly for small values of the parameter $\rho$. 
For $\omega = \rho = 0$, the Kaluza-Klein tower 
exhibits some characteristic features: 
it is the only tower to possess a zero mass state 
while the higher masses are doubly degenerate. 
Indeed, one sees that, for $\rho=0$, when $\omega$ decreases toward zero, 
pairs of adjacent masses in the towers are getting closer and closer 
and take the same value when $\omega$ reaches the value zero.

In Table\re{tableM2a}, Table\re{tableM2c} and Tables\re{tableM2b}-\re{tableM2d}, 
the Kaluza-Klein mass towers 
are similarly presented for a representative choice of bulk mass values,
respectively ${\overline{M}}=0.01$, ${\overline{M}}=0.1$ and ${\overline{M}}=1$ \re{barpara}.  
It appears, as a general rule, 
that the lowest lying mass of the towers $\widetilde{m}_1$ vanishes 
when $\rho=0$  and when the parameter $\omega$ takes 
exactly the value $\omega_{\overline{M}}$
\begin{equation}
\biggl\{\rho=0\ {\rm{and}}\ \omega=\omega_{\overline{M}}= 2\pi R M\biggr\}\ 
\longleftrightarrow m_1=0 \ . 
\label{masszerocondBC1}
\end{equation} 
This agrees with the analogous result for $M=0$ as seen in Table\re{tableM1}.
Moreover, for $\rho=0$ and
for any ${\overline{M}}$ of the order 1, the value of $\widetilde{m}_1$ depends 
almost exactly linearly on the value of the parameter $\omega$ 
from about $\omega=\omega_{\overline{M}}-1$ up to values very close to 
$\omega_{\overline{M}}$ and on the other side from $\omega$ very close 
to $\omega_{\overline{M}}$  
up to about $\omega_{\overline{M}}+1$. 

In general, except for the first mass of the towers in the case $\rho = 0$, 
all the other masses in the towers do not show 
a strong dependence on the value of the $\omega$ parameter.
The fact that 
the first mass of the tower can take values between 0 and about 0.1 TeV, 
and hence can be small
when $\rho$ is not large,
allows one, by a suitable choice of the parameters
$kR$, $M$, $\omega$  
to associate a tower to a particular fermion of the Standard Model 
be it a lepton or a quark and assuming it to be the lowest state
of a Kaluza-Klein tower in a five dimensional warped space.
From the second mass on, the intervals between successive masses 
are generally much larger and more regular. 

In Table\re{tableM3}, Kaluza-Klein towers are presented for the set of boundary 
conditions BC2 (set \re{caseBC2warp}) for the two possible choices of the product
$\epsilon_0\epsilon_R$ of the boundary condition parameters, 
and for each of them, for some values of the reduced bulk mass $\overline{M}$.
In general, the tower masses have a fairly mild 
dependence with respect to the bulk mass, 
with the exception of the first mass in the towers 
when $\epsilon_0\epsilon_R$ is equal to -1, 
in which case, starting from a value of about 0.1 TeV for 
${\overline{M}} = 0$, it falls to less than 
$10^{-10}$ TeV when  ${\overline{M}}$ is equal to one or higher. 
This feature is again of importance in view of practical applications 
of the warp model to fermions, either to the leptons or to the quarks of the Standard Model.
Indeed, by an adequate choice $kR$ and of the bulk mass $M$, 
the first mass of a Kaluza-Klein tower could be made equal 
to the mass of a given lepton, for example the muon, 
leading to identify the tower as associated to that lepton. 
Considering for example $kR=6.3$ and the case of the muon, 
with mass equal to $1.057\, 10^{-4}$ TeV as the lowest mass in a tower, 
the associated Kaluza-Klein tower would result from adopting a value 
around 0.65 for the reduced bulk mass of the 
five-dimensional fermion ${\overline{M_{\mu}}}$ associated to the muon.
A not very different reduced bulk mass
${\overline{M_{e}}}=0.8$ would produce   
the electron of mass equal to $5.11\, 10^{-7}$ TeV as its first mass. 
It is interesting to
remark that bulk fermions with rather close reduced bulk masses (0.65, 0.8)
would lead to the observed fermions with masses in the large ratio
$m_{\mu}/m_e=206.8$. 
It should be noted that
the Kaluza-Klein tower masses associated to either of these 
two leptons would be hardly distinguishable beyond the first mass.
One should also be aware that the Kaluza-Klein towers associated 
to a given fermion would be of a different structure depending on the set 
(BC1 or BC2) of boundary conditions considered.

\subsection{Examples of towers. One metric singularity. 
Semi-local boundary conditions
\label{singtower}}

When there is one metric singularity, the number of arbitrary parameters
increases. Besides $kR$ and $M$, 
the position of the singularity on the strip $[0,2\pi R]$ appears 
as a new parameter
\begin{equation}
s_1=(2\pi R)\,{\overline{s}}_1 \quad,\quad 0\leq {\overline{s}}_1\leq 1 \ .
\label{s1bar}
\end{equation}
which is complemented by the boundary condition parameters.

In order to keep the mass eigenvalues roughly of the order of TeV, we are led to
adapt the value of $kR$ to the value chosen for $\overline{s}_1$. 
Satisfactory choices are
\bea
\overline{s}_1=1&\longleftrightarrow& kR=6.3
       \nonumber\\
\overline{s}_1=0.9&\longleftrightarrow& kR=6.9
       \nonumber\\
\overline{s}_1=0.75&\longleftrightarrow& kR=8.3
       \nonumber\\
\overline{s}_1=0.5&\longleftrightarrow& kR=12.5\ .       
\label{relkrs1}
\eea

There are many possible sets of allowed boundary conditions as seen in the discussion of 
the appendix App.\re{ExempBC}. 
To build our examples, we have limited ourselves to what we call
semi-local boundary conditions: the fields on the left 
and on the right of the singularity are related and, separately, the 
fields at $s=0$ are related to the fields at $s=2\pi R$. Both 
boundary conditions are taken of the form 
BC1 \re{caseBC1warp}, \re{boundarymatrix} and hence are defined 
by four parameters
\bea
BC1 {\rm{\ with\ parameters\ }}
\omega_b,\rho_b && {\rm{at\ the\ edges\ }} 0 {\rm{\ and\ }}2\pi R
    \nonumber\\
BC1 {\rm{\ with\ parameters\ }}
\omega_s,\rho_s && {\rm{on\ both\ sides\ of\ the\ singularity}}\ s_1  .
\label{paras1}
\eea

The conditions for $m_1=0$ are analogous to the conditions 
in the case of no metric singularity \re{masszerocondBC1}
\begin{equation}
\biggl\{\rho_b-\rho_s=0\ {\rm{and\ }}\omega_b-\omega_s= \omega_{\overline{M}}= 2\pi R M\biggr\}\ 
\longleftrightarrow m_1=0\ . 
\label{masszerogen}
\end{equation} 

In Table\re{tableM4a}, for
${\overline{M}}=1$, 
numerical examples of towers
are given
for some arbitrarily chosen positions $s_1$ of the metric singularity 
and some values of the boundary parameters \re{paras1}. 
Similar results, respectively for ${\overline{M}}=0.1$ and ${\overline{M}}=0$, are
presented in Tables
\re{tableM4b} and \re{tableM4c}.

Again in view of applications to leptons and quarks, it should be noted that, when the parameters almost 
satisfy the mass zero conditions \re{masszerogen}, the tower consists 
of a low mass ${\widetilde{m}}_1$ accompanied, as a signature, 
by almost regularly separated doublets of higher masses with ${\widetilde{m}}_2$ 
much larger than ${\widetilde{m}}_1$.

\newpage

\section{Conclusions \label{conclusion}}

In this article, we have extended our previous study 
of the generation of Kaluza-Klein mass towers
for spinor fields propagating in a five dimensional flat space with the fifth dimension compactified 
either on a strip or on a circle.
We have now studied spinor fields propagating in five dimensional 
compactified warp spaces. 

We first considered the case of a warp space without metric singularity. 
We established the specific Dirac equation in the relevant five dimensional warp space for a spinor 
field with an arbitrary bulk mass $M$ and proceeded with the Kaluza-Klein reduction considering the most 
general choice of separation of variables compatible with a SO(3,1) spinor covariance. 
The reducted components of the Dirac fields are found to satisfy a system of
 two coupled equations for which the most general solutions for a four dimensional mass m are given in 
terms of Bessel functions. 

From the requirement of hermiticity of the Dirac operator, we have established 
all the allowed sets of boundary conditions which have to be imposed on the fields.
We found that these boundary conditions belong to two essentially different sets
BC1 \re{caseBC1warp} and BC2 \re{caseBC2warp}, leading to the 
mass equations from which the Kaluza-Klein mass towers can be built.
The same considerations have been extended to the case of warp spaces with an arbitrary number of 
metric singularities. 

In view of the interpretation of the Kaluza-Klein mass eigenstates, specific 
physical
considerations have been made about the possible choices of boundary conditions, about the 
particular situation in which the extradimension strip could be closed to a circle, about the mass 
scale of the model, about the relation between the Kaluza-Klein masses and
the physical masses as observed in a brane and also about the mass state probability densities. 
In particular,
all the parameters with energy dimension 
are scaled to the Plank mass within the only-one-mass-scale hypothesis. 

Finally, illustrative numerical examples of Kaluza-Klein mass towers are given 
when there is no metric singularity for each of the
two sets of boundary conditions, BC1 and BC2. When there is one metric singularity we
have exemplified towers
for some boundary conditions belonging to what we call the semi-local set.
With $kR = 6.3$ or around, it happens that the low lying masses are of the order of TeV, thereby
solving the hierarchy problem without fine tuning.

In the different situations considered, the towers have been established for several choices of 
the basic parameters of the warp model, i.e. the mass reduction parameter $kR=6.3$
(suitably readjusted to the value of $s_1$),
the bulk mass $M$, the position $s_1$ of the singularity on the extradimension if any, as well as of the
parameters defining the boundary conditions.
In general, the 
mass towers are irregularly spaced, and a zero mass state or a small mass 
state exists which depends on the boundary parameters and on the 
value given to the bulk mass $M$. 
This situation allows one, by a suitable choice of the parameters of the model, to associate a mass tower
to any particular fermion of the standard model whose mass would be the smallest in the tower. 
In the assumption that the known leptons and quarks propagate in the bulk under consideration, one would expect
to observe the next low lying masses at high energy colliders, in 
particular at the LHC.

\vspace{5 cm}

\noindent{\bf{Acknowledgment:}}
The authors would like to thank Professor Yves Brihaye for an important suggestion.

\newpage

\begin{appendix}

\section{Least action principle \label{leastact}}

The most general invariant action linear in $\Psi$ and ${\overline{\Psi}}$ 
and of first order in their derivatives is, using \re{covderspin5}, 
\bea
{\cal{A}}
&=&
i\alpha \int{\overline{\Psi}}\gamma^{A}(\overrightarrow{D}_A\Psi)\sqrt{g} d^5x
+i\beta \int({\overline{\Psi}}\,\overleftarrow{D}_A)\gamma^{A}\Psi\sqrt{g} d^5x
  \nonumber\\  
&&\hspace{1 cm}
-m\int{\overline{\Psi}}\Psi\sqrt{g}d^5x\ .
\label{action}
\eea
The underlying Lagrangian is hermitian if 
\be
\beta=-\alpha^*\quad,\quad m=m^*
\label{hermcond}
\ee
and we choose 
\be
\alpha=\frac{\alpha_1+i\alpha_2}{2}\quad,\quad \beta=\frac{-\alpha_1+i\alpha_2}{2}\ .
\label{alphaform}
\ee
Let us note the useful identity
\be
{\overline{\Psi}}\gamma^{A}(\overrightarrow{D}_A\Psi)\sqrt{g}
+({\overline{\Psi}}\,\overleftarrow{D}_A)\gamma^{A}\Psi\sqrt{g} 
=\partial_A\left({\overline{\Psi}}\gamma^{A}\Psi\sqrt{g}\right)\ .
\label{vielbeinid4}
\ee

Requesting then the variation of the action \re{action} 
\bea
\delta{\cal{A}}&=&
 \int{\delta\overline{\Psi}}\biggl(\alpha_1 i\gamma^{A}\overrightarrow{D}_A-m\biggr)\Psi\sqrt{g}\,d^5x
     \nonumber\\
&&-
 \int{\overline{\Psi}}\biggl(\alpha_1 i\overleftarrow{D}_A\gamma^{A}+m\biggr)\delta\Psi\sqrt{g}\,d^5x 
     \nonumber\\
&&+   
  i\frac{\alpha_1+i\alpha_2}{2}\int\partial_A\biggl(\overline{\Psi}\gamma^A\delta\Psi\sqrt{g}  \biggr) d^5x
     \nonumber\\
&&+
  i\frac{-\alpha_1+i\alpha_2}{2}\int\partial_A\biggl(\delta\overline{\Psi}\gamma^A\Psi\sqrt{g}  \biggr) d^5x     
\label{variation}
\eea
to vanish for arbitrary variations 
$\delta\Psi$ and $\delta{\overline{\Psi}}$ of the fields 
$\Psi$ and ${\overline{\Psi}}$, one finds the Dirac equations provided that 
\be
\alpha_1\neq 0 \quad {\rm{,\ conveniently\ normalised\ to:\  }}\ \alpha_1=1\ . 
\label{specalpha}
\ee
They are
\bea
i\gamma^{A}(\overrightarrow{D}_A\Psi)-m\Psi=0
   \nonumber\\
i({\overline{\Psi}}\,\overleftarrow{D}_A)\gamma^{A}+m {\overline{\Psi}}=0  
\label{leastDirac}
\eea
independently of the boundary conditions.
Indeed, since the action is linear in $\Psi$, if it is extremal for two $\delta\Psi$ 
with the same boundary conditions, it is also extremal for their difference 
which is automatically zero at the boundaries. Hence, the field equations are those obtained 
from the usual least action principle i.e. with vanishing variations at the boundaries \re{leastDirac}.

However, further attention has to be devoted to the variation of the action 
arising from the 
boundary terms (third and fourth term in \re{variation}). 
Suppose that there are $N$ 
metric singularities located at the points $s_i,\,i=1,N$ in the $s$ space extending 
from $s_0=0$ to $s_{N+1}=2\pi R$. Denote by 
$\Psi^l(s_i)$ and $\Psi^r(s_i)$  the values of the fields 
at the left and at the right of the metric singularities, and similarly for their variations. 
The boundary relations expressed from the boundary terms in \re{variation} become
\bea
\sum_{i=1}^{N+1} \overline\Psi^l(s_i)\gamma^s\delta\Psi^l(s_i)\sqrt{g(s_i)}
-\sum_{i=0}^{N} \overline\Psi^r(s_i)\gamma^s\delta\Psi^r(s_i)\sqrt{g(s_i)}
    &=&0
\label{BCrel1}\\
 \sum_{i=1}^{N+1} \overline\delta\Psi^l(s_i)\gamma^s\Psi^l(s_i)\sqrt{g(s_i)}
 -\sum_{i=0}^{N} \overline\delta\Psi^r(s_i)\gamma^s\Psi^r(s_i)\sqrt{g(s_i)}
   &=&0 \ .
\label{BCrel2}
\eea
It is natural to suppose that the variations $\delta\Psi$, $\delta{\overline{\psi}}$ 
and the fields 
$\Psi$ and ${\overline{\Psi}}$ belong to the same Hilbert space 
i.e. satisfy the same boundary conditions.  
The relations \re{BCrel1} and \re{BCrel2} then imply boundary conditions 
which turn out to be identical to those obtained in the main 
part of the article from \re{integral} which resulted from the requirement 
of symmetry of the Dirac operator \re{Diracop}.

\newpage
\section{Examples of boundary relations \label{ExempBC}} 

There are many inequivalent sets of boundary conditions related to various choices of
the permutation $P$ in \re{PsiP}. Let us give a few.

\subsection{P=1. Local boundary conditions at the metric singular points. 
Non local conditions at the edges of the $s$-domain \label{semilocalBC}}

With $P=1$, one can obtain
boundary conditions which satisfy the locality conditions (see Sect.\re{locality})
at the singular points but not at the edges. The form of $V^{[2(N+1)]}_P$ 
compatible with this partial-locality is
\be
V^{[2(N+1)]}_P=\left(
                    \matrix{
                     0^{[2]}         &V^{[2]}_1  &0^{[2]}      &0^{[2]}      &\dots     &0^{[2]}      \cr
                     0^{[2]}         &0^{[2]}    &V^{[2]}_2    &0^{[2]}      &\dots     &0^{[2]}      \cr
                     0^{[2]}         &0^{[2]}    &0^{[2]}      &V^{[2]}_3    &\dots     &0^{[2]}      \cr
                     \vdots          &\vdots     &\vdots       &\vdots       &\ddots    &\vdots       \cr
                     0^{[2]}         &0^{[2]}    &0^{[2]}      &0^{[2]}      &\dots     &V^{[2]}_N    \cr
                     V^{[2]}_{N+1}   &0^{[2]}    &0^{[2]}      &0^{[2]}      &\dots     &0^{[2]}
                          } 
               \right) \ .
\label{P1a}
\ee               
Introducing this form of the matrix in the equations \re{relVP}, one finds for all $j$ ($j=1,\dots,N{+}1$)
\bea
  V^{[2]+}_j(i\sigma_2)V^{[2]}_j&=&i\sigma_2
     \label{casea1} \\
  V^{[2]+}_j(\sigma_3)V^{[2]}_j&=&\sigma_3   \ .
     \label{casea2}
\eea
From \re{casea1}, all the $V^{[2]}_j$ are complex symplectic $2\times 2$ matrices restricted by the further
condition \re{casea2}. Their resulting general form is
\be
V^{[2]}_j=e^{i\rho_j}\left(
                     \matrix{ \cosh(\omega_j) & \sinh(\omega_j)\cr
                              \sinh(\omega_j) & \cosh(\omega_j) 
                            }
                     \right)\quad,\quad j=1,\dots,N+1
\label{Vcasea}
\ee
and depends on $2(N{+}1)$ arbitrary parameters. Hence the explicit boundary conditions 
at the metric singularities $j=1,\dots,N$ are 
\begin{equation}
\left(\matrix{F^{[n]l}(s_j)\cr
              G^{[n]l}(s_j)}\right)
              =e^{i\rho_j}\left(
                     \matrix{ \cosh(\omega_j) & \sinh(\omega_j)\cr
                              \sinh(\omega_j) & \cosh(\omega_j) 
                            }\right)
              \left(\matrix{F^{[n]r}(s_j)\cr
              G^{[n]r}(s_j)}\right)
              \ . 
\label{caseBC1warpapp}
\end{equation}
At the edges, the boundary conditions are non local 
\begin{equation}
\left(\matrix{F^{[n]l}(2\pi R)\cr
              G^{[n]l}(2\pi R)}\right)
              =\sqrt[4]{\frac{g(0)}{g(2\pi R)}}\ 
              e^{i\rho_{N+1}}\left(
                     \matrix{ \cosh(\omega_{N+1}) & \sinh(\omega_{N+1})\cr
                              \sinh(\omega_{N+1}) & \cosh(\omega_{N+1}) 
                            }\right)
              \left(\matrix{F^{[n]r}(0)\cr
              G^{[n]r}(0)}\right)
              \ . 
\label{caseBC1warpapp2}
\end{equation}
since they connect the values of the fields at
$s=2\pi R$ to the values of the fields at $s=0$ (a long distance effect). 
When the conditions \re{closurecond} for the closure of the strip to a circle are met, 
these would also be local boundary conditions.

\subsection{ Local boundary conditions both at the metric singular points 
and at the edges of the $s$-domain \label{localBC}}

A way to obtain fully local boundary conditions is to perform the following permutation 
\be
P_2=\left(
    \matrix{ 1      &0      &0      &0      &\dots   &0     &0     &0     \cr
             0      &0      &0      &0      &\dots   &0     &1     &0     \cr
             0      &0      &1      &0      &\dots   &0     &0     &0     \cr
             0      &0      &0      &1      &\dots   &0     &0     &0     \cr
             \vdots &\vdots &\vdots &\vdots &\ddots  &\vdots&\vdots&\vdots\cr
             0      &0      &0      &0      &\dots   &1     &0     &0     \cr
             0      &1      &0      &0      &\dots   &0     &0     &0     \cr
             0      &0      &0      &0      &\dots   &0     &0     &1      
         }
  \right)\ .
\label{P2}         
\ee
and to take $V^{[2(N+1)]}_P$ of the form \re{P1a} with
$V_{N+1}^{[2]}$ diagonal. This leads, on one side to a relation between
$G^{[N]l}(s_{N+1}{=}2\pi R)$ and $F^{[N]l}(s_{N+1}{=}2\pi R)$,  
on the other side to a relation between 
$G^{[N]r}(s_{0}{=}0)$ and $F^{[N]r}(s_{0}{=}0)$.
The conditions for $V_{j}^{[2]}$ ($j=1,\dots,N$) are the same as in
the preceding case \re{casea1}, \re{casea2}, leading to the same 
conditions at each of the singular points \re{Vcasea}. For 
the diagonal $V_{N+1}^{[2]}$,
\bea
V_{N+1}^{[2]+}\sigma_3&=&\sigma_3 V_{N+1}^{[2]}
    \label{convNp1}\\
V_{N+1}^{[2]+}\sigma_3 V_{N+1}^{[2]}&=&\sigma_3 \ .
\label{convNp2}
\eea
Introducing \re{convNp1} in \re{convNp2} on see that $(V_{N+1}^{[2]})^2=1^{[2]}$.      
     
This leads to boundary conditions at the singularities ($s_j$, $j{=}1,\dots,N$) as above
\re{caseBC1warpapp} and to 
\begin{equation}         
\left\{
\begin{array}{rcll}
G^{[n]r}(0)&=&\epsilon_0 F^{[n]r}(0)\quad\quad\ &,\quad  \epsilon_0^2=1
     \cr     
G^{[n]l}(2\pi R)&=&\epsilon_R F^{[n]l}(2\pi R) \quad&,\quad  \epsilon_R^2=1
\end{array}
\right.
\label{caseBC2warpap}
\end{equation}
at the edges.

\subsection{ General boundary conditions for $P=1$ \label{genP}}

When $P=1$, i.e. when the boundary conditions express the values 
at the left of the exceptional points (singularities and edges) in terms of the 
values at the right, the two equations $(j{=}1,2)$ \re{relVP} take the simplified 
form with
$V^{[2(N+1)]}{\equiv}V^{[2(N+1)]}_{P=1}$, $S_{1}^{[2(N+1)]}$ from \re{matrS11}
and $S_{2}^{[2(N+1)]}$ from \re{matrS21c}
\begin{equation}
\left(V^{[2(N+1)]}\right)^+\,S_{j}^{[2(N+1)]}\,V^{[2(N+1)]}=S_{j}^{[2(N+1)]}\ .
\label{relVPone}
\end{equation}
The matrix $V^{[2(N+1)]}$ must be
an element in the intersection of the complex sympletic group
$Sp(2(N{+}1))$ (from the relation for $j=1$) 
and of the pseudo-unitary group $U(N{+}1,N{+}1)$ (from the relation $j=2$).
The dimension of the parameter space can be obtained by 
writing $V^{[2(N+1)]}$ infinitesimally close to the identity 
$1^{[2(N{+}1]}$
\begin{equation}
V^{[2(N+1)]}=1^{[2(N+1)]}+i\eta H^{[2(N+1)]} \quad,\quad \eta\rightarrow 0
\label{infinPone}
\end{equation}
in $2\times 2$ blocks.  
One finds that there are, for $H^{[2(N+1)]}$, $N{+}1$ diagonal 
$2\times 2$ blocks $H_{jj}$ $(j=1,\dots, N{+}1)$, each depending on two real parameters
\begin{equation}
H_{jj}=\left(\matrix{p_j & iq_j\cr
                     iq_j & p_j
             } \right)  \quad,\quad p_j,q_j \ {\rm{real}}
\label{matrixhjj}             
\end{equation}
and $N(N{+}1)$ independent non diagonal $2\times 2$ blocks $H_{jk}$, $j<k=1,\dots, N{+}1$, 
each depending on two complex (four real) parameters $p_{jk}$ and $q_{jk}$
\begin{equation}
H_{jk}=\left(\matrix{p_{jk} & q_{jk}\cr
                     q_{jk} & p_{jk}
             } \right)  \quad,\quad
H_{kj}=\left(\matrix{p^*_{jk} & -q^*_{jk}\cr
                     -q^*_{jk} & p^*_{jk}
             } \right)  \ .      
\label{matrixhjk}             
\end{equation}  
Hence, the set of boundary conditions 
for $P=1$ is indexed by $2(N{+}1)^2$ real parameters.

Let us finally remark that contrary to what happens for the scalar fields where 
the boundary conditions relate the fields and their derivatives, the boundary parameters 
have zero energy dimension \re{barpara} and, hence, there is no need to introduce reduced 
parameters in the spinor case.

\end{appendix}

\newpage 
 
  %%%%%%%%%%%%% Table M1 Mbar = 0 %%%%%%%%%%%%%%%%%%%%%%%%%%%%%%%%%%%%%%%%%

\begin{table}
\caption{
Mass towers for  $M =0$ and for the boundary conditions BC1 
(no metric singularity)
\re{caseBC1warp}, \re{boundarymatrix}.
The towers are symmetric under $\omega\leftrightarrow -\omega$.
The mass eigenvalues ${\widetilde{m}}_i$ are in TeV 
{\label{tableM1}}
}
\vspace{0.5 cm}
\hspace{0.4 cm}
\tiny
{
\begin{tabular}{|c|c|c|c|c|c|c|c|c|c|}
\hline
\multicolumn{10}{|c|}
      {{\rule[-10pt]{0pt}{30pt}}
      {\large{ BC1 (no singularity). Case $M=0$ and $kR=6.3$}}}
        \\ \hline
$\phantom{\bigl[\bigr.}\rho$ &$\omega$
     &$\widetilde{m}_1$&$\widetilde{m}_2$&$\widetilde{m}_3$
     &$\widetilde{m}_4$&$\widetilde{m}_5$&$\widetilde{m}_6$
     &$\widetilde{m}_7$&$\widetilde{m}_8$
      \\ \hline
0&0&
0&0.49362&0.49362&0.98725&0.98725&1.48087&1.48087&1.9740
      \\
&0.01&
0.00078561&0.49283&0.49441&0.98646&0.98803& 1.4801&1.4817&1.9737
      \\
&0.1&
0.00784&0.48578&0.50147&0.97941&0.99509&1.473&1.4887&1.9667
      \\
&1&
0.068017&0.42561&0.56164&0.91923&1.0553&1.4129&1.5489&1.9065
      \\
&5&
0.12235&0.37128&0.61597&0.8649&1.1096&1.3585&1.6032&1.8522
      \\
&100&
0.12341&0.37022&0.61703&0.86384&1.1107&1.3575&1.6043&1.8511     
      \\
\hline      
$\pi/10$
&0&
0.024681&0.46894&0.5183&0.9625&1.01193&1.4562&1.5055&1.9498
      \\
&0.01&
0.024693&0.46893&0.51832&0.96256&1.0119&1.4562&1.5056&2.0004
      \\
&0.1&
0.025858&0.46777&0.51948&0.96139&1.0131&1.455&1.5067&1.9486
      \\
&1&
0.071234&0.42239&0.56486&0.91602&1.0585&1.4096&1.5521&1.9033
      \\
&5&
0.1224&0.37123&0.61603&0.86485&1.1097&1.3585&1.6033&1.8521
      \\
&100&
0.12341&0.37022&0.61703&0.86384&1.1107&1.3575&1.6043&1.8511     
      \\
\hline      
$\pi/4$&0&
0.061703 & 0.43192 & 0.55533 & 0.92555 & 1.049 & 1.4192 & 1.5426 & 1.9128
       \\ 
&0.01&  
0.061707 & 0.43192 & 0.55534 & 0.92554 & 1.049 & 1.4192 & 1.5426 & 1.9128
      \\
&0.1&
0.062094 & 0.43153 & 0.55572 & 0.92516 & 1.0493 & 1.4188 & 1.543 & 1.9124      
      \\
&1&
0.08601 & 0.40762 & 0.57963 & 0.90124 & 1.0733 & 1.3949 & 1.5669 & 1.8885
      \\
&5& 
0.12266 & 0.37097 & 0.61628 & 0.86459 & 1.1099 & 1.3582 & 1.6035 & 1.8518      
      \\
&100&
0.12341 & 0.37022 & 0.61703 & 0.86384 & 1.1107 & 1.3575 & 1.6043 & 1.8511
      \\
\hline      
$\pi/2$&[0,100]&
0.12341&0.37022&0.61703&0.86384&1.1107&1.3575&1.6043&1.8511
       \\ 
\hline      
$3\pi/4$&0&
0.18511&0.30852&0.67874&0.80214&1.1724&1.2958&1.666&1.7894
       \\
&0.1&  
0.18472&0.30891&0.67834&0.80253&1.172&1.2962&1.6656&1.7898
       \\
&1& 
0.1608&0.33282&0.65443&0.82645&1.1481&1.3201&1.6417&1.8137       
       \\
&5& 
0.12415&0.36947&0.61778&0.8631&1.1114&1.3567&1.605&1.8503
       \\
&100& 
0.12341&0.37022&0.61703&0.86384&1.1107&1.3575&1.6043&1.851
       \\
\hline
$9\pi/10$
&0&
0.22213&0.27149&0.71576&0.76512&1.2094&1.2587&1.703&1.7524
\\
&0.1&
0.22095&0.27267&0.71458&0.7663&1.2082&1.2599&1.7018&1.7535
\\
&1&
0.17558&0.31805&0.6692&0.81167&1.1628&1.3053&1.6565&1.7989
\\
&5&
0.12441&0.36921&0.61804&0.86284&1.1117&1.3565&1.6053&1.8501
\\
&100&
0.12341&0.37022&0.61703&0.86384&1.1107&1.3575&1.6043&1.8511     
      \\
\hline
$\pi$&0&
0.24681&0.24682&0.74044&0.74044&1.2341&1.2341&1.7277&1.7277
\\
&0.1&
0.23897&0.25466&0.73259&0.74829&1.2262&1.2419&1.7198&1.7355
\\
&1&
0.1788&0.31483&0.67242&0.80846&1.166&1.3021&1.6597&1.7957
\\
&5&
0.12447&0.36916&0.61809&0.86279&1.1117&1.3564&1.6053&1.85
\\
&100&
0.12341&0.37022&0.61703&0.86384&1.1107&1.3575&1.6043&1.8511     
      \\
\hline
\end{tabular}
   }
\end{table}

  %%%%%%%%%%%%% Table M2a Mbar = 0.01 %%%%%%%%%%%%%%%%%%%%%%%%%%%%%%%%%%%%%%%%%

\vspace{-10 cm}
\begin{table}
\caption{
Mass towers for ${\overline{M}}=0.01$ \re{barpara} 
and for the boundary conditions BC1 
(no metric singularity)
\re{caseBC1warp}, \re{boundarymatrix}.
Here
$\omega_{0.01}=0.3958406\dots$ \re{masszerocondBC1}.
The mass eigenvalues ${\widetilde{m}}_i$  are in TeV 
{\label{tableM2a}}
}\vspace{0.5 cm}
\hspace{-0.5 cm}
\tiny
{
\begin{tabular}{|c|c|c|c|c|c|c|c|c|c|c|}
\hline
\multicolumn{11}{|c|}
{{\rule[-10pt]{0pt}{30pt}}
      {\large{ 
      BC1 (no singularity).
      Case 
      ${\overline{M}}=0.01$ and $kR=6.3$}}}   
           \\ 
           \hline
$\phantom{\bigl[\bigr.}\overline{M}$&$\rho$ &$\omega$
     &$\widetilde{m}_1$&$\widetilde{m}_2$&$\widetilde{m}_3$
     &$\widetilde{m}_4$&$\widetilde{m}_5$&$\widetilde{m}_6$
     &$\widetilde{m}_7$&$\widetilde{m}_8$
      \\ \hline
0.01&0&-100&
0.12486&0.37153&0.61832&0.86511&1.1119&1.3587&1.6055&1.8523
    \\
&&-5&
0.12413&0.37227&0.61758&0.86585&1.1112&1.3595&1.6048&1.8531
    \\
&&-1&
0.08607&0.41059&0.57881&0.90447&1.0722&1.3983&1.5657&1.892
    \\
&&0&
0.03042&0.46598&0.5221&0.96011&1.0152&1.454&1.5085&1.9479
      \\
&&0.01&
0.029692&0.4667&0.52135&0.96084&1.0145&1.4548&1.5078&1.9486
      \\
&&0.05&
0.026728&0.46963&0.51834&0.96377&1.0115&1.4577&1.5048&1.9515      
      \\
&&0.3&
0.0075237&0.48849&0.49891&0.98262&0.99202&1.4765&1.4853&1.9704      
      \\
&&$\omega_{0.01}{-}10^{-3}$&      
7.8549 $10^{-5}$&0.49095&0.49623&0.98406&0.99035&1.4774&1.4843&1.9708
      \\
&&$\omega_{0.01}$&      
0&0.49087&0.49630&0.98398&0.99043&1.47730&1.48435&1.97071      
      \\
&&$\omega_{0.01}{+}10^{-3}$&
7.8546 $10^{-5}$&0.4908&0.49637&0.98391&0.9905&1.4772&1.4844&1.9706
      \\
&&1&
0.04456& 0.44608&0.53983&0.93924&1.0339&1.4326&1.5278&1.926
     \\
&&2&
0.09130&0.3982&0.56746&0.88909&1.0634&1.3812&1.5583&1.8738
     \\
&&5&
0.1204&0.37043&0.61423&0.86409&1.1079&1.3577&1.6015&1.8513
     \\
&&10&
0.12194&0.36891&0.61574&0.86258&1.1094&1.3562&1.603&1.8499
    \\     
&&100&
0.12195&0.3689&0.61575&0.86257&1.1094&1.3562&1.603&1.8498
    \\
\cline{2-11}
&$\pi/10$&-100&
0.12486&0.37153&0.61832&0.86511&1.1119&1.3587&1.6055&1.8523
    \\
&&-5&
0.12417&0.37223&0.61761&0.86582&1.1112&1.3594&1.6048&1.8531
    \\
&&-1&
0.088106&0.4085&0.5809&0.90237&1.0743&1.3961&1.5678&1.8899
    \\ 
&&5 $10^{-3}$&
0.038561&0.45728&0.53078&0.95128&1.024&1.4451&1.5175&1.9389
    \\       
&&0.3&
0.025784&0.46849&0.51891&0.96221&1.0124&1.4559&1.506&1.9495
    \\       
&&$\omega_{0.01}{-}10^{-3}$&
0.024677&0.46877&0.51841&0.96233&1.0121&1.4559&1.5057&1.9495
    \\       
&&$\omega_{0.01}$&
0.024677&0.46876&0.51841&0.96232&1.0121&1.4559&1.5057&1.9495
    \\       
&&$\omega_{0.01}{+}10^{-3}$&
0.024676&0.46876&0.51842&0.96231&1.0121&1.4559&1.5058&1.9495
    \\
&&0.5&
0.025923&0.46665&0.52028&0.96005&1.0141&1.4535&1.5079&1.9471
    \\
&&1&
0.050207&0.44069&0.54522&0.93391&1.0392&1.4273&1.5331&1.9208
    \\
&&5&
0.12048&0.37036&0.6143&0.86402&1.108&1.3576&1.6016&1.8513
    \\           
&&100&
0.12195&0.3689&0.61575&0.86257&1.1094&1.3562&1.603&1.8498
    \\
\cline{2-11}
&$\pi/2$&-100&
0.12486&0.37153&0.61832&0.86511&1.1119&1.3587&1.6055&1.8523
    \\
&&0&
0.12393&0.37068&0.61748&0.86428&1.1111&1.3579&1.6047&1.8515
   \\          
&&$\omega_{0.01}$&      
0.12338&0.37018&0.61699&0.8638&1.1106&1.3574&1.6042&1.851
     \\
&&5&
0.12195&0.3689&0.61575&0.86257&1.1094&1.3562&1.603&1.8498
    \\
&&100&
0.12195&0.3689&0.61575&0.86257&1.1094&1.3562&1.603&1.8498
    \\    
\cline{2-11}
&$\pi$&-100&
0.12486&0.37153&0.61832&0.86511&1.1119&1.3587&1.6055&1.8523
     \\
&&-5&
0.12558&0.3708&0.61905&0.86437&1.1127&1.358&1.6063&1.8516
     \\
&&-1&
0.16353&0.33224&0.65755&0.82548&1.1514&1.3189&1.6451&1.8124
     \\     
&&0&
0.21866&0.27577&0.71308&0.76862&1.2071&1.2619&1.701&1.7552
     \\ 
&&$\omega_{0.01}$&
0.24457&0.24900&0.73739&0.74341&1.23063&1.23740&1.724&1.7313
     \\ 
&&2&
0.15266&0.33865&0.646&0.83258&1.1395&1.3264&1.633&1.8201
    \\
&&5&
0.12349&0.36737&0.61727&0.86106&1.1109&1.3547&1.6045&1.8483
     \\       
&&100&
0.12195&0.3689&0.61575&0.86257&1.1094&1.3562&1.603&1.8498
     \\       
\cline{2-11}
&$3\pi/2$&0&
0.12393&0.37068&0.61748&0.86428&1.1111&1.3579&1.6047&1.8515
     \\
&&$\omega_{0.01}$&
0.12338&0.37018&0.61699&0.8638&1.1106&1.3574&1.6042&1.851
     \\     
&&5&     
0.12195&0.3689&0.61575&0.86257&1.1094&1.3562&1.603&1.8498
     \\
\hline
\end{tabular}
   }
\end{table}

%%%%%%%%%%%%%% table M2c   mbar = 0.1 kr = 6.3

\vspace{-10 cm}
\begin{table}
\caption{
Mass towers for ${\overline{M}}=0.1$ \re{barpara} 
and for the boundary conditions BC1
(no metric singularity) 
\re{caseBC1warp}, \re{boundarymatrix}. Here
$\omega_{0.1}=3.958406\dots$ \re{masszerocondBC1}.
The mass eigenvalues ${\widetilde{m}}_i$ are in TeV 
 \re{boundarymatrix} 
{\label{tableM2c}}
}
\vspace{0.5 cm}
\hspace{-0.2 cm}
\tiny
{
\begin{tabular}{|c|c|c|c|c|c|c|c|c|c|c|}
\hline
\multicolumn{11}{|c|}
{{\rule[-10pt]{0pt}{30pt}}
      {\large{ BC1 (no singularity). Case 
      ${\overline{M}}=0.1$ and $kR=6.3$}}}        
      \\ \hline
$\phantom{\bigl[\bigr.}\overline{M}$&$\rho$ &$\omega$
     &$\widetilde{m}_1$&$\widetilde{m}_2$&$\widetilde{m}_3$
     &$\widetilde{m}_4$&$\widetilde{m}_5$&$\widetilde{m}_6$
     &$\widetilde{m}_7$&$\widetilde{m}_8$
      \\ \hline
0.1&0&-100&
0.13756&0.38327&0.62981&0.87651&1.1232&1.37&1.6168&1.8636
      \\
&&0&
0.13405&0.3982&0.56746&0.88909&1.0634&1.3812&1.5583&1.8738
     \\
&&0.1&
0.13368&0.38757&0.62526&0.88117&1.1184&1.3749&1.6118&1.8686
     \\
&&2&
0.11163&0.41108&0.59908&0.90652&1.0906&1.4013&1.5829&1.8958
     \\ 
&&3.5&
0.035687&0.47886&0.51167&0.97459&1.0017&1.4687&1.4939&1.9622
     \\         
&&$\omega_{0.1}{-}10^{-3}$&      
7.6983 $10^{-5}$&0.4667&0.52135&0.96084&1.0145&1.4548&1.5078&1.9486
      \\
&&$\omega_{0.1}$&      
0&0.46285&0.51711&0.95083&1.0148&1.4412&1.5107&1.9326
      \\
&&$\omega_{0.1}{+}10^{-3}$&
7.6968 $10^{-5}$&0.48849&0.49891&0.98262&0.99202&1.4765&1.4853&1.9704      
      \\
&&5&      
0.09129&0.3982&0.56746&0.88909&1.0634&1.3812&1.5583&1.8738
      \\
&&100&
0.10836&0.35695&0.60413&0.85111&1.098&1.3449&1.5917&1.8386
    \\    
\cline{2-11}
&$\pi/10$&-100&
0.13756&0.38327&0.62981&0.8765&1.1232&1.37&1.6168&1.8636
    \\
&&-5&
0.13754&0.3833&0.62978&0.87653&1.1232&1.37&1.6168&1.8636
    \\
&&0&
0.13422&0.38697&0.6259&0.88051&1.1191&1.3742&1.6125&1.8679
    \\
&&2.5&
0.09733&0.42499&0.58206&0.92123&1.0726&1.4164&1.5644&1.9112
    \\       
&&3.5&
0.04326&0.46584&0.52471&0.96012&1.0162&1.4537&1.5089&1.9471
    \\     
&&$\omega_{0.1}{-}10^{-3}$&      
0.024178&0.4536&0.52631&0.94278&1.0228&1.4338&1.5181&1.9255
      \\
&&$\omega_{0.1}$&      
0.02418&0.45366&0.52627&0.94284&1.0227&1.4338&1.5181&1.9256
      \\
&&$\omega_{0.1}{+}10^{-3}$&
0.024178&0.4536&0.52631&0.94278&1.0228&1.4338&1.5181&1.9255
      \\
&&5.5&      
0.082178&0.3808&0.58249&0.873&1.0776&1.3658&1.572&1.8588
      \\
&&9&      
0.10756&0.35766&0.60346&0.85176&1.0974&1.3455&1.5911&1.8392
      \\
&&15&
0.10836&0.35695&0.60413&0.85111&1.098&1.3449&1.5917&1.8386
      \\
&&100&
0.10836&0.35695&0.60413&0.85111&1.098&1.3449&1.5917&1.8386
    \\
\cline{2-11}
&$\pi/2$&[-100,0]&
0.13755&0.38325&0.62979&0.87648&1.1232&1.37&1.6168&1.8636
   \\          
&&$\omega_{0.1}$&      
0.12055&0.36645&0.61279&0.85929&1.1059&1.3525&1.5991&1.8458
     \\
&&[8,100]&
0.10837&0.35695&0.60414&0.85111&1.098&1.3449&1.5917&1.8386
    \\
\cline{2-11}
&$\pi$&-100&
0.13756&0.38327&0.62981&0.8765&1.1232&1.37&1.6168&1.8636
     \\
&&-5&
0.13759&0.38324&0.62984&0.87647&1.1233&1.37&1.6168&1.8636
     \\
&&0&
0.14106&0.37936&0.6339&0.87225&1.1276&1.3656&1.6213&1.859
     \\ 
&&2&
0.16286&0.35411&0.65892&0.84469&1.154&1.3367&1.6486&1.8292
    \\
&&3.5&
0.22906&0.26911&0.72706&0.75623&1.2218&1.2476&1.7155&1.7403
    \\    
&&$\omega_{0.1}$&
0.22176&0.26629&0.70634&0.76627& 1.1958&1.2629&1.6868&1.7584
     \\ 
&&4.5&
0.18156&0.29665&0.6679&0.79554&1.1584&1.2915&1.6501&1.7867
    \\
&&5&
0.15425&0.31828&0.64339&0.81563&1.1351&1.3109&1.6275&1.8056
     \\       
&&100&
0.10836&0.35695&0.60413&0.85111&1.098&1.3449&1.5917&1.8386
     \\       
\cline{2-11}
&$3\pi/2$&0&
0.13755&0.38325&0.62979&0.87648&1.1232&1.37&1.6168&1.8636
     \\
&&$\omega_{0.1}$&
0.12055&0.36645&0.61279&0.85929&1.1059&1.3525&1.5991&1.8458
     \\     
&&8&     
0.10837&0.35695&0.60414&0.85111&1.098&1.3449&1.5917&1.8386
     \\
\hline
\end{tabular}
   }
\end{table}

\newpage

  %%%%%%%%%%%%% Table M2b Mbar = 1 %%%%%%%%%%%%%%%%%%%%%%%%%%%%%%%%%%%%%%%%%

\vspace{-10 cm}
\begin{table}
\caption{Mass towers 
for ${\overline{M}}=1$ \re{barpara}
as a function of $\omega$ for $\rho=0$ and for the 
boundary conditions BC1 
(no metric singularity) \re{caseBC1warp},\re{boundarymatrix}.
Here $\omega_{1}=39.58406\dots$ \re{masszerocondBC1}).
The masses $m_2$ to $m_8$ are essentially independent of $\rho$. The mass $m_1$ 
is also independent of $\rho$ except for $\omega$ in a range
close to $\omega_{1}$, approximatively in the range [$\omega_{1}{-}10,\omega_{1}{+}10$]. 
In this range,
the variation of $m_1$ as a function of $\rho$ is given in Table\re{tableM2d}.
The mass tower is symmetric under $(\rho)\leftrightarrow (2\pi{-}\rho)$.
The mass eigenvalues ${\widetilde{m_i}}$ are in TeV  
{\label{tableM2b}}
}
\vspace{0.5 cm}
\hspace{-0.1 cm}
\tiny
{
\begin{tabular}{|c|c|c|c|c|c|c|c|c|c|c|}
\hline
\multicolumn{11}{|c|}
 {{\rule[-10pt]{0pt}{30pt}}
      {\large{ BC1 (no singularity). Case 
      ${\overline{M}}=1$ and $kR=6.3$}}}
        \\ \hline
$\phantom{\bigl[\bigr.}\overline{M}$&$\rho$ &$\omega$
     &$\widetilde{m}_1$&$\widetilde{m}_2$&$\widetilde{m}_3$
     &$\widetilde{m}_4$&$\widetilde{m}_5$&$\widetilde{m}_6$
     &$\widetilde{m}_7$&$\widetilde{m}_8$
      \\ \hline
1&0&[-100,5]&
0.24681&0.49363&0.74044&0.98725&1.2341&1.4809&1.7277&1.9745
    \\
&&15&
0.2468&0.49359&0.74039&0.98718&1.234&1.4808&1.7276&1.9744
      \\
&&19&
0.20097&0.41953&0.65138&0.88974&1.1313&1.3746&1.6189&1.8638
      \\
&&20&
0.10356&0.36447&0.61361&0.86141&1.1088&1.3559&1.603&1.85
      \\
&&25&
7.4472 $10^{-4}$&0.35302&0.60692&0.85666&1.1051&1.3529&1.6004&1.8478
      \\
&&30&      
5.0175 $10^{-6}$&
0.35302&0.60692&0.85666&1.1051&1.3529&1.6004&1.8478
      \\
&&39&      
3.8500 $10^{-8}$&0.35302&0.60692&0.85666&1.1051&1.3529&1.6004&1.8478
      \\
&&$\omega_{1}$-$10^{-3}$&      
3.4548 $10^{-13}$&0.35302&0.60692&0.85666&1.1051&1.3529&1.6004&1.8478
      \\
&&$\omega_{1}$&      
0&0.35302&0.60692&0.85666&1.1051&1.3529&1.6004&1.8478
      \\
&&$\omega_{1}$+$10^{-3}$&      
3.4514 $10^{-13}$&0.35302&0.60692&0.85666&1.1051&1.3529&1.6004&1.8478
      \\
&&41&
2.6150 $10^{-10}$&0.35302&0.60692&0.85666&1.1051&1.3529&1.6004&1.8478
     \\
&&50&
3.453 $10^{-10}$&0.35302&0.60692&0.85666&1.1051&1.3529&1.6004&1.8478
     \\
&&[55-200]&
3.4531 $10^{-10}$&0.35302&0.60692&0.85666&1.1051&1.3529&1.6004&1.8478
    \\
\hline
\end{tabular}
   }
\end{table}

  %%%%%%%%%%%%% Table M2d Mbar = 1  m_1 en fonction de omega et rho %%%%%%%%%%%%%%%%%%%%%%%%%%%%%%%%%%%%%%%%%

\newpage

\vspace{-10 cm}

\begin{table}
\caption{
The lowest mass eigenvalue ${\widetilde{m}}_1$
in the towers for ${\overline{M}}=1$ \re{barpara},
as a function of $\rho$ and of
$\omega$ in the range [$\omega_{1}{-}10,\omega_{1}{+}10$],
and for the boundary conditions BC1 
(no 
metric singularity) 
\re{caseBC1warp},\re{boundarymatrix}. Here
$\omega_{1}=39.58406\dots$ \re{masszerocondBC1}).
The mass ${\widetilde{m}}_1$ (in TeV) is symmetric under $(\rho)\leftrightarrow (2\pi{-}\rho)$ 
{\label{tableM2d}}
}
\vspace{0.5 cm}
\hspace{-3.5 cm}
\tiny
{
\begin{tabular}{|c|c|c|c|c|c|c|c|c|c|c|c|c|}
\hline
\multicolumn{13}{|c|}{
      {\rule[-10pt]{0pt}{30pt}}
      {\large{BC1 (no singularity).
      The first mass eigenvalue $\widetilde{m}_1$ for
      ${\overline{M}}=1$ and $kR=6.3$}}}
        \\ 
        \hline
\multicolumn{2}{|c|}{\rule[-3pt]{0pt}{15pt}}
&\multicolumn{11}{|c|}{{\large{$\omega$}}}
        \\
        \cline{3-13}       
\multicolumn{2}{|c|}{\rule[-3pt]{0pt}{15pt}}
&30&36&38&39&$\omega_1{-}10^{-3}$&$\omega_1$&$\omega_1{+}10^{-3}$
&41&42&44&50
       \\
       \hline  
\rule[-3pt]{0pt}{15pt}
&0&5.02 $10^{-6}$& 1.21 $10^{-8}$&1.34 $10^{-9}$&2.74 $10^{-10}$
&3.46 $10^{-13}$&0&3.46 $10^{-13}$&2.62 $10^{-10}$&3.15 $10^{-10}$&3.41 $10^{-10}$&3.45 $10^{-10}$
       \\
       \cline{2-13}                  
\rule[-3pt]{0pt}{15pt}
&$10^{-10}\pi$&5.02 $10^{-6}$& 1.21 $10^{-8}$&1.34 $10^{-9}$&2.74 $10^{-10}$
&3.45 $10^{-13}$&1.08 $10^{-19}$&3.45 $10^{-13}$&2.62 $10^{-10}$&3.14 $10^{-10}$&3.41 $10^{-10}$&3.45 $10^{-10}$
      \\
       \cline{2-13}                  
\rule[-3pt]{0pt}{15pt}
&$10^{-5}\pi$&5.02 $10^{-6}$& 1.21 $10^{-8}$&1.34 $10^{-9}$&2.74 $10^{-10}$
&3.45 $10^{-13}$&1.08 $10^{-14}$&3.45 $10^{-13}$&2.62 $10^{-10}$&3.14 $10^{-10}$&3.41 $10^{-10}$&3.45 $10^{-10}$
      \\
       \cline{2-13}                  
\rule[-3pt]{0pt}{15pt}
&$0.1\pi$&5.02 $10^{-6}$& 1.21 $10^{-8}$&1.36 $10^{-9}$&3.10 $10^{-10}$
&1.08 $10^{-10}$&1.08 $10^{-10}$&1.08 $10^{-10}$&2.67 $10^{-10}$&3.16 $10^{-10}$&3.41 $10^{-10}$&3.45 $10^{-10}$
      \\
       \cline{2-13}                  
\rule[-3pt]{0pt}{15pt}
&$0.2\pi$&5.02 $10^{-6}$& 1.22 $10^{-8}$&1.42 $10^{-9}$&3.96 $10^{-10}$
&2.14 $10^{-10}$&2.13 $10^{-10}$&2.13 $10^{-10}$&2.82 $10^{-10}$&3.21 $10^{-10}$&3.42 $10^{-10}$&3.45 $10^{-10}$
      \\
       \cline{2-13}                  
\rule[-3pt]{0pt}{15pt}
&$0.3\pi$&5.02 $10^{-6}$& 1.22 $10^{-8}$&1.51 $10^{-9}$&5.01 $10^{-10}$
&3.14 $10^{-10}$&3.14 $10^{-10}$&3.13 $10^{-10}$&3.04 $10^{-10}$&3.28 $10^{-10}$&3.43 $10^{-10}$&3.45 $10^{-10}$
      \\
       \cline{2-13}                  
\rule[-3pt]{0pt}{15pt}
&$0.4\pi$&5.02 $10^{-6}$& 1.23 $10^{-8}$&1.61 $10^{-9}$&6.09 $10^{-10}$
&4.06 $10^{-10}$&4.06 $10^{-10}$&4,06 $10^{-10}$&3.29 $10^{-10}$&3.37 $10^{-10}$&3.44 $10^{-10}$&3.45 $10^{-10}$
      \\
       \cline{2-13}                  
\rule[-3pt]{0pt}{15pt}
{\large{$\rho$}}
&$0.5\pi$&5.02 $10^{-6}$& 1.24 $10^{-8}$&1.72 $10^{-9}$&7.09 $10^{-10}$
&4.89 $10^{-10}$&4.88 $10^{-10}$&4,88 $10^{-10}$&3.55 $10^{-10}$&3.47 $10^{-10}$&3.45 $10^{-10}$&3.45 $10^{-10}$
     \\
       \cline{2-13}                  
\rule[-3pt]{0pt}{15pt}
&$0.6\pi$&5.02 $10^{-6}$& 1.25 $10^{-8}$&1.82 $10^{-9}$&7.97 $10^{-10}$
&5.59 $10^{-10}$&5.59 $10^{-10}$&5.58 $10^{-10}$&3.80 $10^{-10}$&3.56 $10^{-10}$&3.47 $10^{-10}$&3.45 $10^{-10}$
      \\
       \cline{2-13}                  
\rule[-3pt]{0pt}{15pt}
&$0.7\pi$&5.02 $10^{-6}$& 1.26 $10^{-8}$&1.91 $10^{-9}$&8.68 $10^{-10}$
&6.16 $10^{-10}$&6.15 $10^{-10}$&6.15 $10^{-10}$&4.00 $10^{-10}$&3.64 $10^{-10}$&3.48 $10^{-10}$&3.45 $10^{-10}$
     \\
       \cline{2-13}                  
\rule[-3pt]{0pt}{15pt}
&$0.8\pi$&5.02 $10^{-6}$& 1.27 $10^{-8}$&1.97 $10^{-9}$&9.21 $10^{-10}$
&6.57 $10^{-10}$&6.57 $10^{-10}$&6.56 $10^{-10}$&4.16 $10^{-10}$&3.71 $10^{-10}$&3.49 $10^{-10}$&3.45 $10^{-10}$
     \\
       \cline{2-13}                  
\rule[-3pt]{0pt}{15pt}
&$0.9\pi$&5.02 $10^{-6}$& 1.28 $10^{-8}$&2.01 $10^{-9}$&9.54 $10^{-10}$
&6.82 $10^{-10}$&6.82 $10^{-10}$&6.82 $10^{-10}$&4.26 $10^{-10}$&3.75 $10^{-10}$&3.49 $10^{-10}$&3.45 $10^{-10}$
     \\
       \cline{2-13}                  
\rule[-3pt]{0pt}{15pt}
&$\pi$&5.02 $10^{-6}$& 1.28 $10^{-8}$&2.03 $10^{-9}$&9.65 $10^{-10}$
&6.91 $10^{-10}$&6.91 $10^{-10}$&6.90 $10^{-10}$&4.29 $10^{-10}$&3.76 $10^{-10}$&3.49 $10^{-10}$&3.45 $10^{-10}$
      \\
      \hline
\end{tabular}
   }
\end{table}

\newpage 
 
\begin{table}
\caption{
Mass towers for the boundary conditions BC2 \re{caseBC2warp} (no metric singularity).
The mass eigenvalues ${\widetilde{m}}_i$ are in TeV 
{\label{tableM3}}
}
\vspace{0.5 cm}
\hspace{-0.1 cm}
\tiny
{
\begin{tabular}{|c|c|c|c|c|c|c|c|c|c|}
\hline
\multicolumn{10}{|c|}
{{\rule[-10pt]{0pt}{30pt}}
      {\large{ BC2 (no singularity).
      Case $kR=6.3$}}}
      \\
      \hline
     \rule[-5pt]{0pt}{15pt}
   $\epsilon_0\epsilon_R$ &${\overline{M}}$
     &$\widetilde{m}_1$&$\widetilde{m}_2$&$\widetilde{m}_3$
     &$\widetilde{m}_4$&$\widetilde{m}_5$&$\widetilde{m}_6$
     &$\widetilde{m}_7$&$\widetilde{m}_8$
      \\ \hline
1
&0&
0.24681&0.49363&0.74044&0.98725&1.2341&1.4809&1.7277&1.9745
      \\
&0.1&
0.25789&0.5053&0.75233&0.99925&1.2461&1.493&1.7398&1.9867
      \\
&0.3&
0.27967&0.52839&0.77592&1.0231&1.2701&1.5171&1.764&2.0109
     \\      
&0.6&
0.31158&0.56244&0.81084&1.0585&1.3058&1.553&1.8001&2.0471
      \\
&0.7&
0.32204&0.57366&0.82237&1.0702&1.3177&1.5649&1.812&2.0591
      \\
&1&
0.35302&0.60692&0.85666&1.1051&1.3529&1.6004&1.8478&2.095
      \\
&1.5&
0.40347&0.66128&0.91289&1.1624&1.411&1.659&1.9067&2.1542
     \\
&2&
0.45279&0.71453&0.96813&1.2189&1.4683&1.7169&1.9651&2.2129
     \\     
&5&
0.73502&1.0187&1.2849&1.544&1.7995&2.0527&2.3045&2.5552
     \\
\hline
-1
&0&
0.12341&0.37022&0.61703&0.86384&1.1107&1.3575&1.6043&1.8511
      \\
&0.1&
0.10836&0.35695&0.60413&0.85111&1.098&1.3449&1.5917&1.8386      
      \\
&0.3&
7.3598 $10^{-2}$&0.32963&0.57794&0.82537&1.0725&1.3195&1.5665&1.8134
      \\
&0.50627&
1.5682 $10^{-2}$&0.30426&0.55445&0.80258&1.0501&1.2974&1.5445&1.7915
      \\
&0.6&      
9.9522 $10^{-4}$&0.31159&0.56246&0.81086&1.0585&1.3059&1.553&1.8001
      \\
&0.7&
2.8062 $10^{-5}$&0.32204&0.57366&0.82237&1.0702&1.3177&1.5649&1.812
      \\
&0.8&
6.8302 $10^{-7}$&0.33244&0.5848&0.83385&1.0819&1.3295&1.5768&1.824
      \\
&0.9&
1.5627 $10^{-8}$&0.34276&0.59589&0.84528&1.0935&1.3412&1.5886&1.8359
      \\
&1&
3.4531 $10^{-10}$&0.35302&0.60692&0.85666&1.1051&1.3529&1.6004&1.8478
      \\
&1.1&
7.4593 $10^{-12}$&0.36321&0.61789&0.86799&1.1166&1.3646&1.6122&1.8596
      \\
&1.2&
1.5857 $10^{-13}$&0.37335&0.62881&0.87928&1.1281&1.3762&1.624&1.8714
      \\
\hline      

\end{tabular}
   }
\end{table}

\newpage

%%%%%%%%%%%%%% table M4a   Une singularite  mbar = 1

\vspace{-10 cm}
\begin{table}
\caption{
Mass towers for ${\overline{M}}=1$ \re{barpara} 
and for the semi-local boundary conditions
(one metric singularity) 
\re{paras1}. Here
$\omega_{1}=2\pi (k R)$ \re{masszerocondBC1}.
The mass eigenvalues ${\widetilde{m}}_i$ are in TeV  
{\label{tableM4a}}
}
\vspace{0.5 cm}
\hspace{-1.5 cm}
\tiny
{
\begin{tabular}{|c|c|c|c|c|c|c||c|c|c|c|c|c|c|c|}
\hline
\multicolumn{15}{|c|}
{{\rule[-10pt]{0pt}{30pt}}
      {\large{ Semi-local BC (one singularity at $s_1=2\pi R{\overline{s}}_1)$. Case 
      ${\overline{M}}=1$ }}}        
      \\ \hline
{\rule[-10pt]{0pt}{30pt}}${\overline{M}}$ & ${\overline{s}}_1$ & $kR$
& $\omega_b$ &$\omega_s$ & $\rho_b$ & $\rho_s$ 
     &$\widetilde{m}_1$&$\widetilde{m}_2$&$\widetilde{m}_3$
     &$\widetilde{m}_4$&$\widetilde{m}_5$&$\widetilde{m}_6$
     &$\widetilde{m}_7$&$\widetilde{m}_8$
      \\ \hline
{\rule[-3pt]{0pt}{10pt}}1&0.9&6.9&2 &0 &0 &0 &
0.28251&0.52712&0.74625&0.96836&1.1863&1.4056&1.6232&1.8416
      \\
{\rule[-3pt]{0pt}{10pt}} & & &15  &0&0 &0&
0.28251&0.52712&0.74625&0.96836&1.1863&1.40558&1.6232 & 1.8416
      \\
{\rule[-3pt]{0pt}{10pt}} & & &20.6  &0&0 &0&
0.28044&0.40672&0.53557&0.75335&0.97614&1.1951&1.4157&1.6346
      \\
{\rule[-3pt]{0pt}{10pt}} & & &20.9  &0&0 &0&
0.27248&0.31334&0.53237&0.75223&0.97546&1.1947&1.4153&1.6342
      \\
{\rule[-3pt]{0pt}{10pt}} & & &21.1  &0&0 &0&     
0.24109&0.29079&0.53166&0.75187&0.97522&1.1945&1.4151&1.6341
      \\
{\rule[-3pt]{0pt}{10pt}} & & &21.5  &0&0 &0&
0.1648&0.28597&0.53108&0.75152&0.97497&1.1943&1.4150&1.6339      
      \\
{\rule[-3pt]{0pt}{10pt}} & & &22  &0&0 &0&
0.10037&0.28481&0.53068&0.75127&0.97478&1.1941&1.4148&1.6338     
      \\ 
{\rule[-3pt]{0pt}{10pt}} & & &25  &0&0 &0&
5.007 $10^{-3}$&0.28481&0.53068&0.75127&0.97478&1.1941&1.4148&1.6338     
      \\ 
{\rule[-3pt]{0pt}{10pt}} & & &28  &0&$0$ &0&
2.493 $10^{-4}$&0.28481&0.53068&0.75127&0.97478&1.1941&1.4148&1.6338      
      \\
{\rule[-3pt]{0pt}{10pt}} & & &35  &0&0 &0&
2.272 $10^{-7}$ &0.28481&0.53068&0.75127&0.97478&1.1941&1.4148&1.6338      
      \\
{\rule[-3pt]{0pt}{10pt}} & & &42  &0&0 &0&
1.538 $10^{-10}$&0.28481&0.53068&0.75127&0.97478&1.1941&1.4148&1.6338      
      \\
{\rule[-3pt]{0pt}{10pt}} & & &$\omega_1{-}10^{-3}$  &0&0 &0&
5.355 $10^{-14}$&0.28481&0.53068&0.75127&0.97478&1.1941&1.4148&1.6338 
    \\
 & & &$\omega_1$  &0&0 &0&
0&0.28481&0.53068&0.75127&0.97478&1.1941&1.4148&1.6338
      \\
{\rule[-3pt]{0pt}{10pt}} & & &$\omega_1{+}10^{-3}$  &0&0 &0&
5.349 $10^{-14}$&0.28481&0.53068&0.75127&0.97478&1.1941&1.4148&1.6338 
    \\
{\rule[-3pt]{0pt}{10pt}} & & &80  &0&0 &0&
5.3520 $10^{-11}$&0.28481&0.53068&0.75127&0.97478&1.1941&1.4148&1.6338 
    \\
\cline{4-15}
{\rule[-3pt]{0pt}{10pt}} & & &$\omega_1{+}2$  &-2&0 &0&
3.492 $10^{-10}$&0.06222&0.60707&0.64552&1.0558&1.0959&1.4971&1.5395
   \\    
{\rule[-3pt]{0pt}{10pt}} & & &$\omega_1$  &0&$\pi$/3 &${-}\pi$/3&
9.2699 $10^{-11}$&0.28481&0.53068&0.75127&0.97478&1.1941&1.4148&1.6338
   \\    
{\rule[-3pt]{0pt}{10pt}} & & &0 &2 &0 &0&
4.5 $10^{-5}$&0.415&0.45234&0.84995&0.8872&1.2846&1.3218&1.7192
    \\ 
{\rule[-3pt]{0pt}{10pt}} & & &2  &0&$\pi/3$ &0&
0.28481&0.53068&0.75127&0.97478&1.1941&1.4148&1.6338&1.8536      
      \\
\hline
{\rule[-3pt]{0pt}{10pt}}1&0.75&8.3&2&0&0&0&
0.2571&0.47971&0.67912&0.88125&1.0795&1.2791&1.4772&1.6758
     \\
{\rule[-3pt]{0pt}{10pt}} & & &$\omega_{1}{-}10^{-3}$ &0& 0 &0&
5.9671 $10^{-16}$&0.25713&0.47971&0.67912&0.88125&1.0795&1.2791&1.4771
     \\
{\rule[-3pt]{0pt}{10pt}} & & &$\omega_{1}$ &0& 0 &0&
0&0.25713&0.47971&0.67912&0.88125&1.0795&1.2791&1.4771
    \\ 
\hline
{\rule[-3pt]{0pt}{10pt}}1&0.5&12.5&2&0&0&0&
0.21972&0.40998&0.5804&0.75315&0.92261&1.0932&1.2624&1.4322
    \\
{\rule[-3pt]{0pt}{10pt}}&&&$\omega_1$&0&0&0&
0&0.21973&0.40997&0.5804&0.75315&0.92261&1.0932&1.2624
    \\
{\rule[-3pt]{0pt}{10pt}}&&&$\omega_1{+}2$&-2&$\pi$/3&$-\pi$/3&
7.0721$10^{-18}$&0.043199&0.46927&0.49822&0.81672&0.84566&1.1586&1.1875
    \\     
\hline
\end{tabular}
   }
\end{table}

%%%%%%%%%%%%%% table M4b   Une singularite  mbar = 0.1

\vspace{-10 cm}
\begin{table}
\caption{
Mass towers for ${\overline{M}}=0.1$ \re{barpara} 
and for the semi-local boundary conditions
(one metric singularity) 
\re{paras1}. Here
$\omega_{0.1}=0.2\pi (kR)$ \re{masszerocondBC1}.
The mass eigenvalues ${\widetilde{m}}_i$ are in TeV 
{\label{tableM4b}}
}
\vspace{0.5 cm}
\hspace{-1.5 cm}
\tiny
{
\begin{tabular}{|c|c|c|c|c|c|c|c|c|c|c|c|c|c|c|}
\hline
\multicolumn{15}{|c|}
{{\rule[-10pt]{0pt}{30pt}}
      {\large{ Semi-local BC (one singularity at $s_1=2\pi R{\overline{s}}_1)$. Case 
      ${\overline{M}}=0.1$ }}}        
      \\ \hline
{\rule[-5pt]{0pt}{15pt}}${\overline{M}}$ & ${\overline{s}}_1$ & $kR$
& $\omega_b$ &$\omega_s$ & $\rho_b$ & $\rho_s$ 
     &$\widetilde{m}_1$&$\widetilde{m}_2$&$\widetilde{m}_3$
     &$\widetilde{m}_4$&$\widetilde{m}_5$&$\widetilde{m}_6$
     &$\widetilde{m}_7$&$\widetilde{m}_8$
      \\ \hline
{\rule[-3pt]{0pt}{10pt}}0.1&0.9&6.9&2 &0 &0 &0 &
0.1108&0.3680&0.5441&1.246&1.413&1.684&1.849&2.121
      \\
{\rule[-3pt]{0pt}{10pt}}& &&$\omega_{0.1}{-}10^{-3}$&0  &0      &0      
&6.711 $10^{-5}$&0.3962&0.4625&0.8241&0.906&1.257&1.347&1.691
 \\
{\rule[-3pt]{0pt}{10pt}}& &&$\omega_{0.1}=4.34...$&0  &0      &0
&0&0.3963&0.4625&0.8242&0.9060&1.257&1.347&1.691
 \\
{\rule[-3pt]{0pt}{10pt}}& &&$\omega_{0.1}{+}10^{-3}$&0  &0      &0
&6.708 $10^{-5}$&0.3962&0.4625&0.8241&0.906&1.257&1.347&1.691
 \\
 \cline{4-15}
{\rule[-3pt]{0pt}{10pt}}& &&0  &$-\omega_{0.1}{-}10^{-3}$&0      &0
&2.181 $10^{-6}$&0.4537&0.4575&0.8892&0.8970&1.324&1.336&1.759
\\
{\rule[-3pt]{0pt}{10pt}}& &&0  &$-\omega_{0.1}$&0      &0
&0&0.4537&0.4575&0.8892&0.8970&1.324&1.336&1.759
\\
{\rule[-3pt]{0pt}{10pt}}   &  & &0  &2  &0      &0
&0.2216&0.2640&0.6553&0.6991&1.091&1.136&1.527&1.573
\\
{\rule[-3pt]{0pt}{10pt}}& &&0  &$\omega_{0.1}$&0      &0
&0.2385&0.2473&0.6717&0.6828&1.106&1.120&1.541 &1.559
\\
 \cline{4-15}
{\rule[-3pt]{0pt}{10pt}}& & &2  &0  &$\pi$/3&0
&0.1189&0.3579&0.5549&0.7954&0.9903&1.233&1.426&1.670
\\
{\rule[-3pt]{0pt}{10pt}}& & &2  &0  &0&$\pi$/3
&0.1189&0.3579&0.5549&0.7954&0.9903&1.233&1.426&1.670
     \\
{\rule[-3pt]{0pt}{10pt}}& &&2  &0  &$\pi$/2&0
&0.1271&0.3481&0.5656&0.7840&1.439&1.657&1.875&2.094       
   \\
 \cline{4-15}
{\rule[-3pt]{0pt}{10pt}}& &&$\omega_{0.1}$&0  &$\pi$/3&0
&0.07014&0.3521&0.5060&0.7855&0.9442&1.221&1.382&1.657
\\
 \hline
\end{tabular}
   }
\end{table}

\newpage

%%%%%%%%%%%%%% table M4c   Une singularite  mbar = 0

\begin{table}
\caption{
Mass towers for $M=0$ and for the semilocal boundary conditions
(one metric singularity) 
\re{paras1}. Here
$\omega_0=0$ \re{masszerocondBC1}.
The mass eigenvalues ${\widetilde{m}}_i$ are in TeV  
{\label{tableM4c}}
}
\vspace{0.5 cm}
\hspace{-1 cm}
\tiny
{
\begin{tabular}{|c|c|c|c|c|c|c|c|c|c|c|c|c|c|c|}
\hline
\multicolumn{15}{|c|}
{{\rule[-10pt]{0pt}{30pt}}
      {\large{ Semi-local BC (one singularity at $s_1=2\pi R{\overline{s}}_1)$. Case 
      $M=0$ }}}        
      \\ \hline
{\rule[-10pt]{0pt}{30pt}}${\overline{M}}$ & ${\overline{s}}_1$ & $kR$
& $\omega_b$ &$\omega_s$ & $\rho_b$ & $\rho_s$ 
     &$\widetilde{m}_1$&$\widetilde{m}_2$&$\widetilde{m}_3$
     &$\widetilde{m}_4$&$\widetilde{m}_5$&$\widetilde{m}_6$
     &$\widetilde{m}_7$&$\widetilde{m}_8$
      \\ \hline
{\rule[-3pt]{0pt}{10pt}}0&0.9&6.9&-2 &0 &0 &0 &
0.09061&0.3467&0.5279&0.7841&0.9653&1.221&1.403&1.659
      \\
{\rule[-3pt]{0pt}{10pt}}   &  & &$-10^{-3}$  & 0&0      &0
&6.966 $10^{-5}$&0.4373&0.4374&0.8746&0.8747&1.312&1.312&1.749
\\
{\rule[-3pt]{0pt}{10pt}}   &  & &0  &0  &0      &0
&0&0.4373&0.4373&0.8746&0.8746&1.312&1.312&1.749
\\
{\rule[-3pt]{0pt}{10pt}}   &  & &$10^{-3}$  & 0&0      &0
&6.961 $10^{-5}$&0.4373&0.4374&0.8746&0.8747&1.3119&1.3121&1.7492
\\
{\rule[-3pt]{0pt}{10pt}} & & &2 &0 &0 &0 &
0.09061&0.3467&0.5279&0.7841&0.9653&1.221&1.403&1.659
      \\
\cline{4-15}
{\rule[-3pt]{0pt}{10pt}}& & &$2{+}10^{-3}$  &2  &$\pi$/3&$\pi/3$
&1.849 $10^{-5}$&0.4346&0.4401&0.8691&0.8803&1.304&1.320&1.738
\\
 \hline
\end{tabular}
   }
\end{table}

\end{document}